\documentclass[useAMS,usenatbib]{mn2e}
\usepackage{epsfig}

 \font\sevenrm=cmr7 scaled 1000
\def\gsim{\;\lower4pt\hbox{${\buildrel\displaystyle >\over\sim}$}\;}
\def\lsim{\;\lower4pt\hbox{${\buildrel\displaystyle <\over\sim}$}\;}
\def\grls{\;\lower4pt\hbox{${\buildrel\displaystyle >\over <}$}\;}
\title[MHD EECC solutions]
{Envelope Expansion with Core Collapse \\
II. Quasi-Spherical Self-Similar Solutions\\
for an Isothermal Magnetofluid }
\author[C. Yu \& Y.-Q. Lou]{Cong Yu$^{1,5}$
\thanks{A substantial part of this research publication was 
carried out at Physics Department and Tsinghua Center for 
Astrophysics (THCA) of Tsinghua University; this research 
publication serves as a partial fulfillment of PhD thesis 
requirement for C.Y.    }
\thanks{E-mail: yccit@yahoo.com.cn; louyq@mail.tsinghua.edu.cn
and lou@oddjob.uchicago.edu}
and Yu-Qing Lou$^{2,3,4}$\\
$^{1}$National Astronomical Observatories/Yunnan Astronomical
Observatory, Chinese Academy of Sciences, Kunming, 650011, China;\\
$^{2}$Physics Department and Tsinghua Center for Astrophysics
(THCA), Tsinghua University, Beijing, 100084, China;\\
$^{3}$Department of Astronomy and Astrophysics, the University
of Chicago, 5640 South Ellis Avenue, Chicago, IL 60637, USA;\\
$^{4}$National Astronomical Observatories, Chinese Academy
of Sciences, A20, Datun Road, Beijing 100012, China; \\
$^{5}$Graduate School of the Chinese Academy of Sciences, Beijing, China.}
\begin{document}

\date{Accepted 2005 ?? ??. Received 2005 ?? ??;
in original form 2005 April ??}

\pagerange{\pageref{firstpage}--\pageref{lastpage}} \pubyear{2004}

\maketitle

\label{firstpage}

\begin{abstract}
We investigate self-similar magnetohydrodynamic (MHD) processes in an
isothermal self-gravitating fluid with a quasi-spherical symmetry and
extend the envelope expansion with core collapse (EECC) solutions of
Lou \& Shen by incorporating a random magnetic field. Magnetized
expansion-wave collapse solutions (MEWCS) can be readily constructed
as a special case. This inside-out MHD collapse occurs at the
magnetosonic speed and magnetized EECC solutions are obtained
systematically. Stagnation surfaces of EECC solutions that seperate
core collapse and envelope expansion propagate at constant speeds
either sub-magnetosonically or super-magnetosonically. These
similarity MHD solutions show various behaviours, such as radial
inflow or contraction, core collapse, oscillations and outflow or
wind as well as shocks. For solutions to go across the magnetosonic
line smoothly, we carefully analyze topological properties of
magnetosonic critical points.
Without crossing the magnetosonic critical line, continuous
spectra of magnetized EECC and envelope contraction with core
collapse (ECCC) solutions are readily obtained. Crossing the 
magnetosonic line twice analytically, there exists an infinite 
number of discrete magnetized EECC and ECCC solutions. One or 
more sub-magnetosonic stagnation surfaces associated with these 
discrete solutions travel outward at constant yet different speeds 
in a self-similar manner. In addition to the EECC shock solution 
which could change the central accretion rate, the magnetic field 
can also affect the core accretion rate. As the magnetic parameter 
$\lambda$ increases, the core accretion rate appropriate for the 
MHD EWCS becomes larger. Under the frozen-in approximation, 
magnetic fields in the envelope expansion portion would scale as 
$B\propto r^{-1}$, while in the core collapse portion they would 
scale as $B\propto r^{-1/2}$. We discuss several astrophysical 
applications of EECC similarity solutions to the formation process 
of proto-planetary nebulae connecting the AGB phase and the 
planetary nebula, to supernova remnants, collapse of magnetized 
molecular cloud, to $\textrm{H}\ {\textrm{\small{II}}}$ clouds 
surrounding massive OB stars and to a certain evolution phase 
of galaxy clusters.

\end{abstract}


\section{Introduction}

The process of gravitational collapse is fundamental in various
cosmological and astrophysical contexts, and is certainly one
important phase for dynamical processes of star formation. Over
decades, observational and theoretical studies have been combined
together to outline a plausible scenario for such gravitational
collapses in gaseous molecular clouds, taking account of 
geometrical factors and other main physical considerations (e.g., 
Bodenheimer \& Sweigart 1968). Self-similar solutions under the 
simplifying assumption of spherical symmetry have long been 
pursued in the past \citep{larson69a,penston69a,shu77,hunter77,
hunter86,whitworth85,fosterchevalier93}. Independently,
\citet{penston69a,penston69b} and \citet{larson69a,larson69b} started
a simple analytical description of spherically symmetric isothermal
collapse under the self-gravity that reproduced essential features of 
more detailed numerical simulations. In a seminal paper that introduced 
the current paradigm for an `inside-out collapse' in low-mass star 
formation, \citet{shu77} derived the so-called self-similar expansion 
wave collapse solution (EWCS). Subsequently, \citet{hunter77} carried 
out a more comprehensive search and found an infinite number of discrete
solutions in the `complete solution space'. \citet{whitworth85} have
considerably expanded these solutions into a two-parameter continuum
by allowing for weak discontinuities across the sonic critical line.

For molecular clouds that are initially near a marginally stable
equilibrium, the Larson-Penston (LP) solution provides a sensible
approximation to the condition near the centre of the point mass
(i.e., a protostar) formation epoch. \citet{shu77} derived the
EWCS that has a free-fall core collapse with a mass density
profile of $\rho\propto r^{-3/2}$, a radial infall speed profile
$u\propto r^{-1/2}$, a constant central mass accretion rate
at $r\rightarrow 0^{+}$, and a static envelope of a singular
isothermal sphere (SIS) with the collapsing front moving outward
at the constant sound speed $a$, where $r$ is the radial distance
from the centre. This EWCS scenario of an inside-out collapse for
star formation has been advocated by \cite{sal87} and compared
with observations \citep{als,zhou93,choi95,sa99,harvey01}.
\citet{fosterchevalier93} studied the gravitational collapse of
an isothermal sphere by hydrodynamical simulations and recovered
the LP solution in the central region where a core forms and the
self-similar solution of \citet{shu77} when the ratio of initial
outer cloud radius to core radius is $\gsim 20$.

Molecular gas clouds from which many stars come into being are
structurally complex and highly filamentary in many places. The
origin of such complex structures is not well understood at
present. It may be partially produced by regular and random
magnetic fields on different scales (e.g., Balick \& Frank 2002).
The formation of filaments and sheets may be necessary to the
fragmentation of molecular clouds into clumps of order of the
final stellar masses, although the relation between the initial
cloud mass and the mass of the resulting stars remains unclear (e.g., 
Hartmann 1998). Due to the formation of filamentary structures in 
numerical simulations \citep{porter94,klessen00,ostriker01} and 
the appearance of numerous filaments in observations (e.g., 
Falgarone et al. 2001), analyses on processes of filament formation 
and evolution have also been pursued in the literature 
\citep{kawachi98,hennebelle03,tp03,shadmehri05}.

In view of the ubiquitous presence of magnetic fields in
the interstellar medium and their nontrivial dynamic and
diagnostic roles in astrophysical plasma systems, dynamical
collapses of magnetized clouds have generated considerable
interests. In some cases, magnetic fields may not be
dynamically important, yet can offer valuable diagnostic
information through cyclotron or synchrotron emissions.
\citet{chiueh} carried out an investigation
of dynamical collapse in nonrotating magnetized gas clumps of grossly
spherical shape. They focussed on the magnetic pressure effect with 
the magnetic tension effect being ignored. Effects of magnetic fields 
have also been extensively investigated in various contexts
\citep{contopoulos98,krasnopolsky02,hennebelle03,tp03,shadmehri05}.
In terms of theoretical formulation and astrophysical applications, 
we note that magnetic field can make a tenuous collisionless gas to 
behave more fluid like as evidenced by fusion plasma experiments 
and solar wind observations. As a result, we could study dilute 
magnetized plasmas on large scales using the MHD description.

Thermal radio emissions from planetary nebulae have been observed
\citep{kwok82,kwok85,kwok93} and might perhaps indicate the presence
of magnetic fields. By extensive observations of various morphologies,
it is most likely that magnetic field should influence the shapes and
shaping of PNe (e.g. Balick \& Frank 2002). Although not a PN, the
remnant morphology of a triple ring system of the supernova SN 1987A
may also hint at important roles of magnetic field as modelled through 
numerical simulations \citep{tanaka02}. Within the collapsed core of 
a PN, magnetic dynamo processes may also operate for some time to 
generate and sustain a strong magnetic field (e.g., $\sim 10^4-10^9$G)
associated with a proto-white dwarf \citep{blackman01}.

%

The role of magnetic field in our model analysis is characterized
by one important parameter $\lambda$ as defined by equation (\ref{intro}).
Different systems of astrophysical objects involve various values of
$\lambda$. The estimated $\lambda$ parameter for the Crab Nebula is 
of the order of a magnitude $\sim 10^{5}$. For a galaxy cluster, the 
estimated $\lambda$ is approximately 0.2. For star forming regions,
the estimated $\lambda$ is typically 0.01. For planetary nebulae, the 
typical $\lambda$ is approximately $\sim 0.003$.

%



In addition to collapses and flows with an isothermal equation 
of state, gravitational collapses with polytropic or logotropic 
equations of state have also been investigated by several authors
\citep{cheng78,goldreich80,bouquet85,suto88,mclaughlin97,
kawachi98,scalo98,maeda02,harada03,shadmehri05}. With a polytropic
equation of state, the gravitational bound system have different
temperatures at different radii. It is also possible to include
effects of radiative losses in the energy equation (e.g., Boily
\& Lynden-Bell 1995; Murakami, Nishihara \& Hanawa 2004).
Stability of self-similar gravitational
collapses was also considered by several authors
\citep{ori88,hanawa97,hanawa99,hanawa00,laigoldreich00,lai00}.
\citet{laigoldreich00} studied the growth of nonspherical
perturbations in the collapse of a self-gravitating spherical gas
cloud. They found that nonspherical perturbations damp in the
subsonic region and grow in the supersonic region. They mentioned
potential applications to core collapse of supernova explosions
\citep{cheng78,goldreich80,yahil83}, where asymmetric density
perturbations may lead to asymmetric shock propagation and breakout,
giving rise to asymmetry in the explosion and a kick to a new-borne
neutron star. Global stability analysis with a polytropic equation 
of state was also carried out by \citet{lai00}.
Slow rotation effects on gravitational collapses has been considered 
by \citet{tsc}. As an important extension of Shu (1977), the general
relativistic analysis of collapse or accretion onto a black hole has 
been recently conducted by Cai \& Shu (2005).

Recently, \citet{loushen} examined self-similar spherical isothermal
collapses and derived new solutions in the semi-complete solution
space; this is in reference to the `complete solution space' of 
\citet{hunter77}. The gas flow at large $r$, together with the two 
distinct asymptotic solution behaviours near the origin, lead to 
an infinite number of discrete self-similar solutions that are 
analytically smooth across the sonic critical line. Those solutions 
characterized by envelope expansion with core collapse (EECC) are of 
particular interest. In various contexts of astrophysical systems, 
it would be natural and important to include effects of magnetic 
fields in a collapsing cloud. Our main motivation here is to study 
this generalized class of EECC solution in the presence of a random
magnetic field. Meanwhile, we broaden and enrich all the known 
similarity solution structures in the most general manner.

In the presence of magnetic field, the exact spherical symmetry for a
magnetized flow no longer exists in the rigorous sense. Nevertheless,
morphological observations of supernova remnants, such as the Crab
Nebula, show that gross quasi-elliptical structures do exist on large
scales. With these empirical considerations in mind, we develop 
an MHD formalism for a quasi-spherical symmetry for large-scale 
magnetized flows. The close analogy between the mass conservation 
equation and the magnetic induction equation gives rise to a special 
relation connecting the gas mass density, the magnetic field component 
transverse to the radial flow direction and the radius $r$. With this 
relation, magnetic field can be readily incorporated into the EECC 
framework \citep{loushen}. This paper is to investigate the magnetic 
field effects on the EECC solutions under the approximation of
a quasi-spherical symmetry.

Besides these shock-free globally smooth solutions, \citet{th95}
constructed a self-similar shock solution describing a situation
in which a central thermal or kinetic energy release initiates an
outgoing shock during a protostellar collapse of a low-mass star.
Such a shock solution, matched with a static SIS envelope, can 
have a central free-fall collapse, yet with a central mass 
accretion rate lower than that predicted by the EWCS. \citet{th95}
also constructed a shock expansion solution with a finite core density
matched with a static SIS envelope; this solution has been recently
generalized and subsumed into a family of `champagne flow' shock
solutions by \citep{shu02} to model expansions of H {\sevenrm II}
regions surrounding massive OB stars after the passage of an initial 
ionization front. To be more general, \citet{shenlou04} studied EECC 
shock solutions allowing outflows or inflows far away from the central 
region. In terms of modelling proto-stellar systems, this flexibility 
can accommodate a variety of real possibilities. Most recently, Bian 
\& Lou (2005) explored the parameter space more systematically to 
construct various isothermal shock solutions, including for example, 
twin shocks, and so forth.

This paper is structured as follows. The basic MHD equations and
the self-similar transformation are presented in \S\ 2. In \S\ 3, we
provide a comprehensive analysis of the resulting nonlinear MHD ODEs.
In \S\ 4, we solve the nonlinear MHD ODEs numerically with the help
of necessary analytical analysis given in \S\ 3. In \S\ 5, we give 
the physical interpretations of our results. 

\section[]{PHYSICAL ASSUMPTIONS AND
\\ \ \ \quad  BASIC MODEL FORMULATION}

We adopt the isothermal approximation for the gas under consideration. 
By the quasi-spherical symmetry of a magnetized flow system, the basic 
MHD equations in spherical polar coordinates $(r,\theta,\phi)$ are:
\begin{equation} \label{continuity}
\frac{\partial \rho}{\partial t}+\frac{1}{r^{2}}
\frac{\partial (r^{2}\rho u)}{\partial r}=0\ ,
\end{equation}
\begin{equation}\label{muequivalent1}
\frac{\partial M}{\partial r}=4\pi\rho r^{2}\
\end{equation}
for the mass conservation, where $u$ is the bulk radial flow speed,
$\rho$ is the gas mass density and $M(r,t)$ is the enclosed mass 
within $r$ at time $t$. By the definition of $M(r,t)$, we have the 
equivalent of equation (\ref{continuity}), namely
\begin{equation}\label{muequivalent2}
\frac{\partial M}{\partial t}+u\frac{\partial M}{\partial r}=0\ .
\end{equation}
The radial momentum equation is simply
\begin{equation} \label{momentum}
\frac{\partial u}{\partial t}+u\frac{\partial u}{\partial r}
=-\frac{a^{2}}{\rho}\frac{\partial \rho }{\partial r}-
\frac{GM}{r^{2}}-\frac{1}{8\pi\rho}\frac{\partial}{\partial r}
<B^{2}_{\perp}>-\frac{<B^{2}_{\perp}>}{4\pi\rho r}
\end{equation}
where $-\partial\Phi/\partial r\equiv -GM(r,t)/r^2$ with $\Phi(r,t)$
being the gravitational potential and $B_{\perp}$ is the magnitude of 
the magnetic field transverse to the radial direction. The Poisson 
equation relating the gas mass density $\rho$ and the gravitational 
potential $\Phi$ is automatically satisfied. Here in equation 
(\ref{momentum}), we keep the magnetic tension force term that was 
ignored in equation (3) of \citet{chiueh}.
%

The MHD energy conservation takes the form of
\[
\frac{\partial}{\partial t}\bigg\{\frac{\rho u^2}{2}
+a^2\rho\bigg[\ln\bigg(\frac{\rho}{\rho_{c}}\bigg)-1\bigg]
-\frac{1}{8\pi G}\bigg(\frac{\partial\Phi}{\partial r}\bigg)^{2}
+\frac{\mathbf{B}^2}{8\pi}\bigg\}
\]
\[
\qquad\qquad+\frac{1}{r^2}\frac{\partial}{\partial r}
\bigg\{r^{2}u\rho\bigg[\frac{u^2}{2}
+a^2\ln\bigg(\frac{\rho}{\rho_c}\bigg)\bigg]\bigg\}
\]
\[
\qquad\qquad+\frac{1}{r^2}\frac{\partial}{\partial r}
(r^2u\rho\Phi)+\frac{1}{r^2}\frac{\partial}{\partial r}
\bigg(\frac{r^2\Phi}{4\pi G}\frac{\partial^2\Phi}
{\partial r\partial t}\bigg)
\]
\begin{equation}
\qquad\qquad+\nabla\cdot\bigg[\frac{\mathbf{B}}{4\pi}
\times(\mathbf{u}\times\mathbf{B})\bigg]=0\ ,
\end{equation} 
which generalizes equation (4) of Lou \& Shen (2004) by 
including the magnetic field $\mathbf{B}$; the magnetic 
energy density and the Poynting flux density can be 
readily identified, respectively. We derive this MHD 
energy conservation equation when all magnetic field 
components are included. 

To reach the above MHD energy conservation equation, we have
taken the electrical conductivity to be infinite, such that
the magnetic induction equation takes the form of
\begin{equation} \label{induction2}
\frac{\partial \mathbf{B}}{\partial t}
=\nabla\times(\mathbf{u}\times\mathbf{B})\ .
\end{equation}
The three component forms of this magnetic induction equation are
\[
\frac{\partial B_{r}}{\partial t}
=\frac{1}{r\sin\theta}\frac{\partial}{\partial\theta}
\bigg[\sin\theta(u_{r}B_{\theta}-u_{\theta}B_{r})\bigg]
\]
\begin{equation}\label{radial}
\qquad\qquad\qquad\qquad
-\frac{1}{r\sin\theta}\frac{\partial}{\partial\phi}
(u_{\phi}B_{r}-u_{r}B_{\phi})\ ,
\end{equation}
\[
\frac{\partial B_{\theta}}{\partial t}
=\frac{1}{r\sin\theta}\frac{\partial}{\partial\phi}
(u_{\theta}B_{\phi}-u_{\phi}B_{\theta})
\]
\begin{equation}\label{theta}
\qquad\qquad\qquad\qquad
-\frac{1}{r}\frac{\partial}{\partial r}
\bigg[r(u_{r}B_{\theta}-u_{\theta}B_{r})\bigg]\ ,
\end{equation}
\[
\frac{\partial B_{\phi}}{\partial t}
=\frac{1}{r}\frac{\partial}{\partial r}
\bigg[r(u_{\phi}B_{r}-u_{r}B_{\phi})\bigg]
\]
\begin{equation}\label{phi}
\qquad\qquad\qquad\qquad
-\frac{1}{r}\frac{\partial}{\partial\theta}
(u_{\theta}B_{\phi}-u_{\phi}B_{\theta})\ ,
\end{equation}
where $u_r\equiv u$ in our formalism.
We just consider radial bulk flow $u_r\neq 0$
by setting $u_{\theta}=u_{\phi}=0$, equation
(\ref{radial}) then becomes
%
\begin{equation}\label{radial6}
\frac{\partial B_{r}}{\partial t}
=u_{r}(\nabla_{\perp}\cdot\mathbf{B}_{\perp})
+(\mathbf{B}_{\perp}\cdot\nabla)u_{r}\ .
\end{equation}
Using the divergence-free condition of magnetic field
$\nabla\cdot\mathbf{B}=0$, equation (\ref{radial6}) becomes
\begin{equation}
\frac{1}{2}\frac{D}{D t}[(r^{2}B_{r})^{2}]
=r^{4}B_{r}(\mathbf{B}_{\perp}\cdot\nabla)u_{r}\ ,
\end{equation}
where $D/Dt\equiv\partial/\partial t+u_{r}\partial/\partial r$.
The $\theta-$ and $\phi-$components of the magnetic induction
equations (\ref{theta}) and (\ref{phi}), under the simplifying
approximation of $u_{\theta}=0$ and $u_{\phi}=0$, become
\begin{equation}\label{theta4}
\frac{D }{D t}\ln\bigg(\frac{B_{\theta}}{r}\bigg)=
-\frac{1}{r^{2}}\frac{\partial}{\partial r}(r^{2}u_{r})\ ,
\end{equation}
\begin{equation}\label{phi4}
\frac{D }{D t}\ln\bigg(\frac{B_{\phi}}{r}\bigg)=
-\frac{1}{r^{2}}\frac{\partial}{\partial r}(r^{2}u_{r})\ .
\end{equation}
The continuity equation of mass conservation
takes a similar form in parallel, namely
\begin{equation}\label{rho}
\frac{D }{D t}\ln\rho = -\frac{1}{r^{2}}
\frac{\partial}{\partial r}(r^{2} u_{r})\ .
\end{equation}
From the above three equations 
(\ref{theta4})$-$(\ref{rho}), we obtain
\begin{equation}\label{rhoBr}
\frac{B_{\theta}}{\rho r}=\textrm{const}\ ,\quad\quad
\frac{B_{\phi}}{\rho r}=\textrm{const}\ ,
\quad\quad\frac{B_{\perp}}
{\rho r}=\textrm{const}\ ,
\end{equation}
where $B_{\perp}=(B^{2}_{\theta}+B^{2}_{\phi})^{1/2}$ is 
the magnitude of the magnetic field component transverse 
to the radial component.

Based on the last of integrals (\ref{rhoBr}),
we introduce a simple relation
\begin{equation}\label{intro}
B^{2}_{\perp}=16\pi^{2}\lambda G\rho^{2}r^{2}
\end{equation}
with the magnetic parameter $\lambda$ being a 
proportional constant. In comparison with the work of 
\citet{chiueh}, it is apparent that
\[
\lambda=\frac{\beta(x)}{\alpha^{2}x^{2}}\ ,
\]
where $\beta(x)$ is a dimensionless function introduced by
\citet{chiueh} and $\alpha$ is our reduced density. Note 
that the ratio of the Alfv\'en speed $v_A$ to the isothermal 
sound speed $a$ is given by
\[
\qquad \frac{v_{A}}{a}=\bigg(\frac{\beta}{\alpha}\bigg)^{1/2}
\qquad\hbox{ and }\qquad
v_{A}\equiv\frac{B_{\perp}}{(4\pi\rho)^{1/2}}\ .
\]
Similarity solutions with $x$ ranging from
$x\rightarrow+\infty$ to $x\rightarrow0^{+}$
are referred to as the semi-complete solutions
(Shu 1977; Whitworth \& Summers 1985;
Lou \& Shen 2004; Shen \& Lou 2004).
Similarity solutions with $x$ ranging from $x\rightarrow-\infty$
to $x\rightarrow+\infty$ are referred to as complete solutions
\citep{hunter77,hunter86}. The perspectives of complete
\citep{hunter77,hunter86} and semi-complete
\citep{shu77,whitworth85,loushen} similarity solutions are both
valid with appropriate physical interpretations and with the
proper specification of an initial moment. For the invariance
under the time
reversal transformation of self-gravitational magnetofluid equations
$(\ref{continuity})-(\ref{momentum})$, namely, $t\rightarrow-t$,
$\rho\rightarrow\rho$, $u\rightarrow-u$, the semi-complete and complete
similarity solutions are in fact closely related to each other. Complete
Hunter type solutions can be decomposed into two branches of similarity
solutions in the semi-complete space (Lou \& Shen 2004). Futhermore, the
time reversal transformation can also lead to different interpretations
and application of self-similar solutions. In this paper, we focus on
the semi-complete solutions. For ideal MHD equations with an infinite
electric conductivity, the magnetic induction equation (\ref{induction2})
is invariant under $t\rightarrow-t$ and $\vec v\rightarrow -\vec v$
transformation. In fact, the corresponding transformation of either
$\vec B\rightarrow\vec B$ or $\vec B\rightarrow -\vec B$ is allowed.
If we take the magnetic diffusion term into account, these invariant
properties are no longer valid.

We now introduce the self-similar transformation:
\[
\rho(r,t)=\frac{\alpha(x)}{4\pi G t^{2}}\ ,
\]
\[
M(r,t)=\frac{a^{3}t}{G}m(x)\ ,
\]
\begin{equation}
u(r,t)=a v(x)\ ,
\end{equation}
\[
\Phi(r,t)=a^{2}\phi(x)\ ,
\]
\[
B_{\perp}=\frac{a}{\sqrt{G}t}b(x)\ ,
\]
and immediately get
\[
b(x)=\sqrt{\lambda}\alpha x\ ,
\]
where $x\equiv r/(at)$ is the independent self-similar variable;
the dimensionless $\alpha(x)$, $m(x)$, $v(x)$, $\phi(x)$ and
$b(x)$ are the reduced dependent variables for the gas mass 
density, the enclosed mass, the radial flow speed, the 
gravitational potential and the transverse magnetic field, 
respectively. They are all functions of $x$ only. Note that 
$b(x)$ is related to $\beta(x)$ of \citet{chiueh} by 
$b^{2}\equiv\beta$. After a direct substitution of the above 
transformations into the nonlinear MHD partial differential 
equations, we obtain the following nonlinear ordinary 
differential equations (ODEs).

It follows that equations (\ref{muequivalent1})
and (\ref{muequivalent2}) reduce to
\begin{equation}\label{muequiv1}
m^{'}=x^{2}\alpha\ ,
\end{equation}
\begin{equation}\label{muequiv2}
(v-x)m^{'}+m=0\ ,
\end{equation}
where the prime denotes the derivative with respect to $x$. Combining
these two equations, we immediately obtain the following two equations
\begin{equation}\label{mpositive}
m=(x-v)x^{2}\alpha\ ,
\end{equation}
\begin{equation}\label{alphav}
\frac{d}{dx}[x^{2}\alpha(x-v)]=x^{2}\alpha\ .
\end{equation}
By equation (\ref{mpositive}), the physical condition of a
positive mass $m(x)>0$ is transformed into the condition 
of $x-v>0$. Therefore, solutions of $v(x)$ must lie to the 
upper-right of the straight line $x-v=0$ in the plane 
$-v(x)$ versus $x$.

The continuity equation of mass conservation becomes
\begin{equation}\label{dimensionlesscontinue}
v^{'}+(v-x)\frac{\alpha^{'}}{\alpha}=\frac{2}{x}(x-v)\ 
\end{equation}
and the radial momentum equation becomes
\begin{equation}\label{dimensionlessmomentum}
(v-x)v^{'}+(1+\lambda\alpha x^{2})
\frac{\alpha^{'}}{\alpha}=-\alpha(x-v)-2\lambda x\alpha\ .
\end{equation}
Rearranging these two equations above, we arrive at
\[
\bigg[(x-v)^{2}-(1+\lambda\alpha x^{2})\bigg]v^{'}
\]
\begin{equation}\label{mainv}
\qquad\qquad\qquad\qquad=(x-v)\bigg[\alpha(x-v)-\frac{2}{x}\bigg]\ 
\end{equation}
and
\[
\bigg[(x-v)^{2}-(1+\lambda\alpha
x^{2})\bigg]\frac{\alpha^{'}}{\alpha}
\]
\begin{equation}\label{mainalpha}
\qquad\qquad\qquad\qquad=(x-v)\bigg[\alpha-\frac{2}{x}(x-v)\bigg]
+2\lambda x\alpha\ .
\end{equation}
Note that by setting $\lambda=0$ for the absence of magnetic field,
we readily recover the formulation of Lou \& Shen (2004). For
smooth and regular solutions crossing the magnetosonic line,
the critical conditions are
\begin{equation}
(x-v)\bigg[\alpha(x-v)-\frac{2}{x}\bigg]=0\ ,
\end{equation}
\begin{equation}\label{msonic}
(x-v)^{2}-(1+\lambda\alpha x^{2})=0\ ,
\end{equation}
\begin{equation}
(x-v)\bigg[\alpha-\frac{2}{x}(x-v)\bigg]+2\lambda x\alpha=0\
\end{equation}
in order to determine finite first derivatives $v'$ and $\alpha'$,
respectively. In fact, only two of the above three requirements 
are independent. For $x-v\neq 0$, the above three equations are 
equivalent to the following three equations
\begin{equation}\label{critical1}
x-v=\frac{2}{\alpha x}\ ,
\end{equation}
\begin{equation}\label{critical2}
\alpha^{2}x^{2}+\lambda\alpha^{3}x^{4}=4\ ,
\end{equation}
\begin{equation}\label{critical3}
(x-v)+2\lambda x=(x-v)^{3}\ .
\end{equation}
For $\lambda=0$, equation (\ref{critical3}) becomes $(x-v)=\pm 1$ 
and equation (\ref{critical2}) reduces to $\alpha=2/x$. Given the 
positiveness of $x-v>0$, we then come to the familiar critical 
condition $x-v=1$ \citep{loushen}. Equation ({\ref{msonic}) for 
$(x-v)^{2}-(1+\lambda\alpha x^{2})=0$ is equivalent to the condition 
$(x-v)^{2}=(a^{2}+v^{2}_{A})/a^{2}$ in dimensional form, directly 
related to the magnetosonic condition.

\section{Analysis of Nonlinear ODEs}

\subsection[]{Special Analytical Solutions}


With $\phi^{'}=mx^{-2}$, an exact global analytical solution 
of nonlinear ODEs (\ref{mainv}) and (\ref{mainalpha}) appears 
in the form of 
\begin{equation}\label{specialone}
v=0,\quad\alpha=\frac{2}{x^{2}},\quad m=2x,\quad
\frac{d\phi}{dx}=\frac{2}{x},
\quad b=\frac{2{\lambda}^{1/2}}{x}\ ,
\end{equation}
which describes a magnetostatic equilibrium
for a quasi-spherical system
\citep{ebert55,bonnor56,chandrasekhar57,shu77}.
In physical dimensions, we simply have
\[
\quad\rho=\frac{a^{2}}{2\pi G r^{2}}\ ,
\qquad M=\frac{2a^{2}r}{G}\ ,
\qquad B_{\perp}^2=\frac{4\lambda a^4}{Gr^2}\ .
\]
In this solution, the magnetic pressure gradient force density and
the magnetic tension force density cancel each other (i.e., a nearly
force-free magnetic field), that is, their combined effects on the
entire system are absent. To be more specific but in dimensionless
form, the sum of the magnetic pressure gradient and tension force
densities is simply
\[
\qquad -\lambda(\alpha^{'}x^{2}+2\alpha x)=0\ .
\]
%
%
Another exact global analytical similarity solution of nonlinear
ODEs (\ref{mainv}) and (\ref{mainalpha}) is
\[
v=\frac{2}{3}x\ ,\qquad\alpha=\frac{2}{(3+18\lambda )}\ ,
\qquad m=\frac{2x^{3}}{3(3+18\lambda)}\ ,
\]
\begin{equation}\label{specialtwo}
\frac{d\phi}{dx}=\frac{2x}{3(3+18\lambda)}\ ,
\qquad b=\frac{2\lambda^{1/2}x}{(3+18\lambda )}\ .
\end{equation}
In physical dimensions, we simply have
\[
\rho=\frac{1}{2(3+18\lambda)\pi G t^{2}}\ ,
\qquad M=\frac{2 r^{3}}{3(3+18\lambda)G t^{2}}\ ,
\]
\[
B_{\perp}^2={4\lambda r^2\over G(3+18\lambda)^2t^4}\ .
\]
Within a finite radius $r$, the transverse magnetic field strength
$|B_{\perp}|$ scales as $t^{-2}$. This solution passes the
magnetosonic critical line at $x=3(1+6\lambda)^{1/2}$ as would be
evident in our later discussion about the MHD generalization of
the Larson-Penston-type solutions. In the absence of magnetic
field with $\lambda=0$, this solution passes the sonic critical
line at $x=3$ \citep{whitworth85}. Such solution corresponds to a
nonrelativistic Einstein-de Sitter expansion with magnetic field.
Here, the `cosmic mass density' is $\rho=1/[(6+36\lambda)\pi Gt^{2}]$
and the relevant `Hubble constant' is $H=2/[3t(1+6\lambda)^{1/2}]$
that decreases with time $t$.

Another exact global solution is the singular solution
\begin{equation}\label{specialthree}
x-v=\frac{2}{\alpha x}\quad\hbox{ and }
\quad (x-v)^{2}-(1+\lambda\alpha x^{2})=0\ ,
\end{equation}
which actually specifies the magnetosonic critical line.
This set of equations for the  magnetosonic critical
line shall be discussed in more details presently.

The second condition of equation
(\ref{specialthree}) is equivalent to
\[
a(x-v)=(v_{A}^{2}+a^{2})^{1/2}\ ;
\]
the right-hand side is the magnetosonic speed and the left-hand
side is the wave front speed relative the local flow speed.
When the Doppler-shifted flow speed coincides with the local 
magnetosonic speed, the singularity would arise. The situation 
appears to be similar to the discussion of the `nozzle condition' 
in the steady solar wind \citep{parker63}, or to the discussion 
of the critical accretion rate in a steady spherical accretion 
\citep{bondi52}.


\subsection[]{Results for Crossing the Magnetosonic Critical Line}

The coupled nonlinear MHD equations are singular along the magnetosonic 
critical line. Because of this, we cannot directly obtain similarity 
solutions that go across the magnetosonic critical line smoothly by a 
simple-minded numerical integration backward from $x\rightarrow+\infty$. 
To obtain solutions satisfying magnetosonic critical conditions 
$(\ref{critical1})-(\ref{critical3})$, we need to first derive 
non-trivial first derivatives of $v$ and $\alpha$ by applying the 
L'H\^{o}spital rule along the magnetosonic critical line. Higher-order 
derivatives of $v$ and $\alpha$ across the magnetosonic critical line 
can also be determined accordingly \citep{hunter86,whitworth85}. It is 
straightforward to derive a quadratic equation for $z\equiv v^{'}$, namely
\begin{equation}\label{criticalderivative}
2\bigg[(v-x)-\frac{x\lambda}{(v-x)^2}\bigg]z^2
+2(x-v)z-\frac{2v}{x^2}=0
\end{equation}
(see Appendix B for a detailed derivation).

Solving this quadratic equation, together with equation
(\ref{critical1}), (\ref{critical3}) and
(\ref{dimensionlesscontinue}), we obtain the first derivative $v'$
along the magnetosonic critical line. Once the first derivative
$v'$ is known, we immediately obtain the first derivative $\alpha'$
using equations (\ref{dimensionlesscontinue}) or
(\ref{dimensionlessmomentum}). By this procedure, the problem of
crossing the magnetosonic critical line smoothly and analytically
has been completely solved by means of an effective numerical scheme.
For quadratic equation (\ref{criticalderivative}), analytical
expansions could be found once the coefficients of this equation
can be given explicitly. We have solved the magnetosonic critical
line equation as described later in this section, that is, we know 
the the value of $v$ for a given value of $x$ along the magnetosonic 
critical line. Thus analytical eigensolutions of equation 
(\ref{criticalderivative}) can be given, as all the coefficients of 
the equation (\ref{criticalderivative}) can be expressed in terms 
of $x$. However, such explicit eigensolutions of equation 
(\ref{criticalderivative}) would be very complex and it would be 
more convenient to derive the two eigensolutions numerically.

In comparison with equations $(16)-(18)$ of \citet{chiueh}, we find
that their results about the magnetosonic critical point is a very
special case of ours. By setting $v=0$, we get similar results as
those of \citet{chiueh}. Their equations $(16)-(18)$ are results
about the zero point $x_{c}=(1+2\lambda)^{1/2}$. Expressing our
results related to the zero point in terms of their similar
notations, we have
\[
\lambda=\frac{x_{c}^{2}-1}{2}\ ,
\quad\alpha_{c}=\frac{2}{x_{c}^{2}}\ ,
\quad\beta_{c}=b_{c}^{2}=\frac{2(x_{c}^{2}-1)}{x_{c}^{2}}\ ,
\]
\begin{equation}\label{ewcsv}
v_{c}=0\ ,\quad
v_{c1}^{'}=\frac{2x_{c}^{2}}{3x_{c}^{2}-1}\ ,
\quad v_{c2}^{'}=0\ ,
\end{equation}
\begin{equation}\label{ewcsa}
\alpha_{c1}^{'}=-\frac{4(2x_{c}^{2}-1)}{x_{c}^{3}(3x_{c}^{2}-1)},\quad
\alpha_{c2}^{'}=-\frac{4}{x_{c}^{3}}\ ,
\end{equation}
\begin{equation}
\beta_{c1}^{'}
=2\lambda\alpha_{c}\alpha_{c1}^{'}x_{c}^{2}+2\lambda\alpha_{c}^{2}x_{c}
=-\frac{8\lambda(x_{c}^{2}-1)}{x_{c}^{3}(3x_{c}^{2}-1)}\ ,
\end{equation}
\begin{equation}
\beta_{c2}^{'}
=2\lambda\alpha_{c}\alpha_{c2}^{'}x_{c}^{2}+2\lambda\alpha_{c}^{2}x_{c}
=-\frac{8\lambda}{x_{c}^{3}}\ .
\end{equation}
Note that $\lambda\geqslant0$ and thus $x_{c}\geqslant1$
\citep{chiueh}. The derivative of $b=\lambda^{1/2}\alpha x$ 
can be calculated once results of $\alpha$ are known. The 
key difference lies in the fact that our momentum equation 
contains the magnetic tension force term that was ignored 
in their formulation. These results are crucial for 
constructing MHD expansion-wave collapse solutions (MEWCSs) 
\citep{shu77,chiueh}.\footnote{We note in passing that 
equations $(1)-(3)$ and $(8)-(10)$ of Chiueh \& Chou (1994) 
are somehow inconsistent. The magnetosonic critical 
condition in their analysis is correct.}

By equation (\ref{criticalderivative}), there are two types 
of eigensolutions for $z\equiv v'$ at each point along the 
magnetosonic critical line. When the two roots of equation 
(\ref{criticalderivative}) are of opposite signs, type 1 and 
type 2 eigensolutions are referred to those with negative and 
positive roots of the equation (\ref{criticalderivative}), 
respectively. When the two roots of equation 
(\ref{criticalderivative}) are of the same sign, type 1 and 2 
eigensolutions are those with smaller and larger absolute 
values respectively,
following the nomenclature of Shu (1977) and Lou \& Shen (2004).
In Lou \& Shen (2004), type 1 corresponds to $dv/dx=1-1/x_{\ast}$
and type 2 corresponds to $dv/dx=1/x_{\ast}$. In the open range of
$0<x<2$, type 1 and type 2 are exactly defined as such. When $x>2$
(relevant to the LP-type solution), $dv/dx$ of type 1 has a larger
absolute value, while $dv/dx$ of type 2 has a smaller absolute
value, that is, their magnitudes reverse for nodal points.
In our current definition, no magnitude reversal would happen.
When the point is a nodal point, type 1 and type 2 solutions
remain always the smaller and larger ones respectively for the
absolute value of $dv/dx$. To summarize, our definition is not
defined by the explicit expressions of $dv/dx$ (e.g.,
$1-1/x_{\ast}$ and $1/x_{\ast}$), but by their magnitudes and
signs. It is easier to keep in mind their relevant physical
properties.

As a necessary consistent check, we note that when $\lambda=0$ 
and $x-v=1$, the familiar results of Shu (1977) are readily 
recovered, namely
\begin{equation}\label{shueqn}
-2z^{2}+2z-\frac{2(x-1)}{x^{2}}=0\ .
\end{equation}
The two eigensolutions of this quadratic equation are simply
\[
z_1=\frac{1}{x_{\ast}},
\qquad\qquad\qquad
z_2=1-\frac{1}{x_{\ast}}
\]
[see the Appendix of Shu (1977)].

Expanding equation (\ref{critical3}) and regrouping terms,
we obtain the following cubic equation in terms of $v(x)$
\begin{equation}\label{cubiceqn}
v^3+a_2v^2+a_1v+a_0=0\
\end{equation}
describing the magnetosonic critical 
line, where the three coefficients are
$
a_2=-3x ,
$
$
a_1=3x^2-1\
$
and
$
a_0=2\lambda x+x-x^3\ .
$
This equation can be solved for $v$ in terms of $x$. Let 
$z_{1}$, $z_{2}$, $z_{3}$ be the three roots of cubic 
equation (\ref{cubiceqn}) in terms of $x$. By the 
properties of a cubic equation, it follows that
\begin{equation}
z_{1}+z_{2}+z_{3}=-a_{2}\ ,
\end{equation}
\begin{equation}
z_{1}z_{2}+z_{2}z_{3}+z_{3}z_{1}=a_{1}\ ,
\end{equation}
\begin{equation}
z_{1}z_{2}z_{3}=-a_{0}\ .
\end{equation}
Following the standard procedure of solving a cubic algebraic 
equation, we then introduce two dimensionless parameter $q$ 
and $r$ such that
\begin{equation}
q=\frac{a_{1}}{3}-\frac{a_{2}^{2}}{9}\ ,
\end{equation}
\begin{equation}
r=\frac{(a_{1}a_{2}-3a_{0})}{6}-\frac{a_{2}^{3}}{27}\ .
\end{equation}
It then follows that
\[
\qquad
q^{3}+r^{2}=\frac{1}{3}\bigg(\frac{a_{1}}{6}\bigg)^{2}(4a_{1}-a_{2}^{2})
\]
\begin{equation}
\qquad\qquad\qquad\qquad+a_{0}\bigg(\frac{a_{2}}{3}\bigg)^{3}
+\frac{a_{0}^{2}}{4}-\frac{a_{0}a_{1}a_{2}}{6}\ .
\end{equation}
For $q^{3}+r^{2}>0$, the cubic equation possesses one real root for
$v$ and a pair of complex conjugate roots for $v$. For $q^{3}+r^{2}=0$,
all three roots of $v$ are real and at least two of them are equal.
For $q^{3}+r^{2}<0$, all three roots of $v$ are real. In cubic
equation (\ref{cubiceqn}), $v$ is the unknown for a given $x$.
In solving for $v$, we express the three roots $z_i$ as
\begin{equation}\label{realroot}
z_{1}=(s_{1}+s_{2})-{a_{2}}/{3}\ ,
\end{equation}
\begin{equation}
z_{2, 3}=-\frac{(s_{1}+s_{2})}{2}-\frac{a_{2}}{3}
\pm\frac{i\sqrt{3}}{2}(s_{1}-s_{2})\ ,
\end{equation}
where the two parameters $s_{1}$ and $s_{2}$ are defined by
\[
s_{1,2}\equiv [r\pm(q^{3}+r^{2})^{1/2}]^{1/3}\ .
\]

Specific to the current problem, we simply have
\begin{equation}
q=-{1}/{3}\ ,
\end{equation}
\begin{equation}
r=-x\lambda
\end{equation}
and
\begin{equation}
q^{3}+r^{2}=x^{2}\lambda^{2}-{1}/{27}\ ,
\end{equation}
when
\begin{equation}
x\geqslant {1}/({3\sqrt{3}\lambda})\ ;
\end{equation}
the only real root is the one given by equation (\ref{realroot}).
Given an explicit expression of this root, we can analyze the nodal
or saddle properties along the magnetosonic critical line (e.g.,
Jordan \& Smith 1977). For $x\leqslant(3\sqrt{3}\lambda)^{-1}$, we
need to identify the suitable root lying in the required parameter
regime to satisfy $x-v>0$. This is somewhat more involved than the
case of $x\geqslant(3\sqrt{3}\lambda)^{-1}$.

For the case of $x\leqslant(3\sqrt{3}\lambda)^{-1}$, we adopt an
alternative solution procedure by introducing a transitional variable
\begin{equation}
\theta\equiv\arccos\bigg(\frac{-r}{\sqrt{-q^{3}}}\bigg)
=\arccos(3\sqrt{3}x\lambda)\ ,
\end{equation}
where $\theta$ lies in the range $0\leqslant\theta\leqslant\pi/2$.
The three roots of the cubic equation can be expressed more
compactly and neatly in terms of this $\theta$ parameter
\citep{press86}, namely
\[
z_{1}=-2\sqrt{-q}\cos\bigg(\frac{\theta}{3}\bigg)-\frac{a_{2}}{3}
\]
\begin{equation}\label{joinsmall}
\qquad=-\frac{2\sqrt{3}}{3}
\cos\bigg(\frac{\theta}{3}\bigg)+\frac{x}{3}\ ,
\end{equation}
\[
z_{2,3}=-2\sqrt{-q}\cos\bigg(\frac{\theta\pm2\pi}{3}\bigg)
-\frac{a_{2}}{3}
\]
\begin{equation}
\qquad=-\frac{2\sqrt{3}}{3}
\cos\bigg(\frac{\theta\pm2\pi}{3}\bigg)+\frac{x}{3}\ .
\end{equation}
Since $\theta/3$ falls in the range of
$0\leqslant\theta\leqslant\pi/6$, it follows that $\cos(\theta/3)$ is
positive while $\cos(\theta/3+2\pi/3)$ and $\cos(\theta/3-2\pi/3)$
are both negative. Therefore, $z_{2}$ and $z_{3}$ must be positive,
while $z_{1}$ may be either positive or negative. The zero point
(i.e., $-v=0$) of the magnetosonic critical line is at
\begin{equation}
x=(1+2\lambda)^{1/2}\ .
\end{equation}
The other special location is
\begin{equation}
x={1}/(3\sqrt{3}\lambda)
\end{equation}
that separates $-v$ regimes for one root (for larger $|x|$)
and three real roots (for smaller $|x|$). By equating these
two special locations, we have
\begin{equation}
\lambda_{*}={1}/{6}\ .
\end{equation}
\begin{figure}
\includegraphics[width=3in]{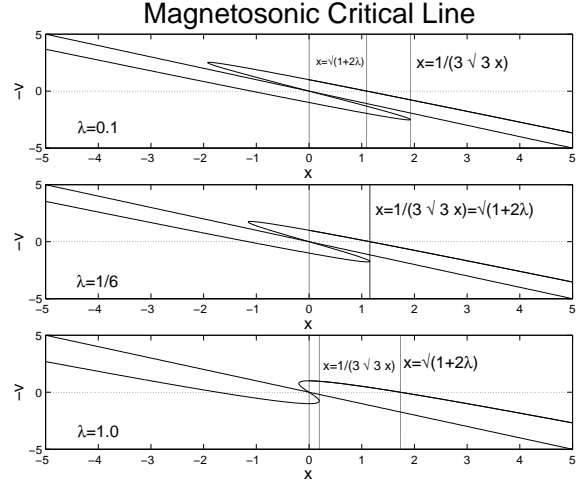}
\caption{\label{comparison} The magnetosonic critical lines for
$\lambda=0.1,\ 1/6,\ 1.0$ in upper, middle, lower panels, respectively;
the vertical coordinate is the negative reduced radial speed
$-v\equiv -u/a$ and the horizontal coordinate is the independent
similarity variable $x\equiv r/(at)$. In the upper panel for
$\lambda=0.1$, the zero point is to the left of the turning point. In
the middle panel for $\lambda=1/6$, the zero point and the turning are
at the same location. In the lower panel for $\lambda=1$, the zero
point is to the right of the turning point. This diagram is consistent
with our analytical analysis. The diagonal straight line from the
upper-left to the bottom-right in each panel is $x-v=0$. }

\end{figure}
We refer to the point with abscissa $x=1/(3\sqrt{3}\lambda)$ in
the fourth quadrant as the turning point along the magnetosonic
critical line. In Figure 1, we see that for $\lambda=1/6$, the
zero point and the turning point of the curve in the fourth
quadrant have the same abscissa $x$ as expected. For $\lambda<1/6$,
the abscissa of the zero point is smaller than that of the turning
point in the fourth quadrant, while for $\lambda>1/6$, the abscissa
of the zero point is larger than that of the turning point of the
curve in the fourth quadrant.
From the three panels in Fig. 1, we conclude that the
magnetosonic critical lines are connected smoothly by
the two solutions (\ref{joinsmall}) and (\ref{realroot})
at $x=(3\sqrt{3}\lambda)^{-1}$.
A quick summary for the magnetosonic critical
line in terms of $-v$ is as follows: when
$x\leqslant(3\sqrt{3}\lambda)^{-1}$,
$z_{1}=-(2/\sqrt{3})\cos({\theta}/{3})+x/3$; when
$x\geqslant(3\sqrt{3}\lambda)^{-1}$,
$z_{1}=s_{1}+s_{2}+x/3$.
Once the magnetosonic critical line in terms of $v(x)$ is
obtained, the corresponding magnetosonic critical line
for $\alpha(x)$ can be readily determined by equation
(\ref{critical1}).
In Figure 2, we explore variations of the magnetosonic
critical lines for different values of $\lambda$. By
the time reversal symmetry of the magnetosonic critical
line, we plot them in the entire plane. Our focus is
mainly on the first and fourth quadrants.
\begin{figure}
 \includegraphics[width=3in]{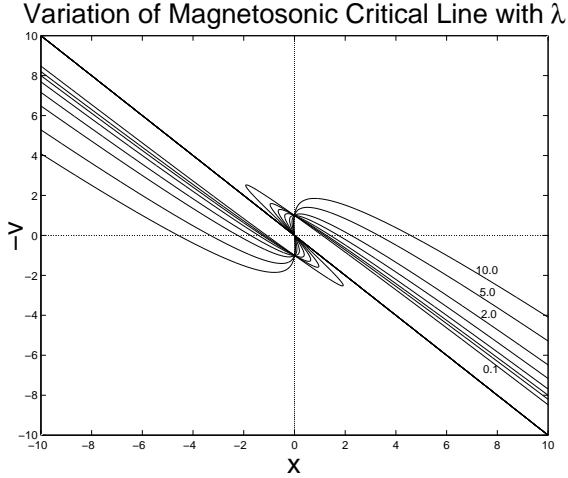}
 \caption{\label{sonicline} We choose $\lambda=0.1,\ 0.2,\ 0.3,\
 0.5,\ 1.0,\ 2.0,\ 5.0,\ 10.0$ (from the lower part to the top
 part in order in the first and fourth quadrants); the vertical
 coordinate is the negative reduced radial speed $-v\equiv -u/a$
 and the horizontal coordinate is the independent self-similarity
 variable $x\equiv r/(at)$. It is clear in this figure that in the
 first and fourth quadrants, all the magnetosonic critical lines
 lie above the straight line $-v=\sqrt{3}/3-x$. This information
 is essential for our analysis of the nodal and saddle points in
 Appendix B. The diagonal straight line from the upper-left to 
 the bottom-right is $x-v=0$.  }
\end{figure}

\subsection[]{Asymptotic Solutions of Nonlinear ODEs}

Solution behaviours of nonlinear ODEs {(\ref{mainv})} and
{(\ref{mainalpha})} at infinity ($x\rightarrow +\infty$) 
show effects of magnetic field (i.e., $\lambda\neq 0$) in 
the higher order terms, as seen clearly from the following 
expressions of asymptotic expansions
\[
v=V+\frac{2-A}{x}+\frac{V}{x^{2}}
\]
\begin{equation}\label{infinity11}
\quad+\frac{(A/6-1)(A-2)+2V^{2}/3+A(2-A)\lambda/3}{x^{3}}+\cdots\ ,
\label{vinfty}
\end{equation}
%
\[
\alpha=\frac{A}{x^{2}}+\frac{A(2-A)}{2x^{4}}+\frac{A(4-A)V}{3x^{5}}+
\]
\[
\frac{A(A-3)(A/2-1)-(A-6)A V^{2}/4+(2-A)A^{2}\lambda/4}{x^{6}}
\]
\begin{equation}\label{infinity22}
\hspace{2em}+\cdots\ ,
\label{alphainfty}
\end{equation}
\[
\beta=\lambda\alpha^{2} x^{2}=\frac{A^2\lambda}{x^2}+\frac{A^2
(2-A)\lambda}{x^4}+\frac{A^2(2-A)^2\lambda}{4 x^6}
\]
\begin{equation}\label{infinity33}
\hspace{2em}+\cdots\ ,
\end{equation}
where $V$, $A$ and $\lambda$ are three integration constants
referred to as the speed, density and magnetic field parameters,
respectively. The magnetic field parameter $\lambda$ comes into
play fairly late in the expansion of large $x$. In the reduced
velocity expansion, magnetic fields first enter in the $x^{-3}$
term. In the reduced density expansion, magnetic fields first
enter in the $x^{-6}$ term. So the MHD solution behaviours at 
large $x$ are only slightly modified by the presence of magnetic 
field. The magnitude of transverse magnetic field $B_{\perp}$ at 
large distances scales as
$$
B_{\perp}^{2}=\frac{\lambda a^4 A^2}{G r^2}\ .
$$
We may get back to nonlinear ODEs {(\ref{mainv})} and
(\ref{mainalpha}) to qualitatively see the leading expansion
behaviours of the solutions (\ref{vinfty}) and (\ref{alphainfty}).
In the limit of $x\rightarrow+\infty$, the leading term of $v$ is 
the constant speed parameter $V$ and the leading term of $\alpha$ 
is $Ax^{-2}$. By substituting the leading terms into the following 
expressions for various relevant forces, we obtain the pressure 
gradient force in the limit of $x\rightarrow+\infty$ as
\[
-{\alpha^{'}}/{\alpha}={2}/{x}\ ,
\]
the gravitational force in the limit of $x\rightarrow+\infty$ as
\[
-\alpha(x-v)=-\frac{A}{x}\ ,
\]
and the magnetic pressure and tension forces 
together in the limit of $x\rightarrow+\infty$ as
\[
-\lambda(\alpha^{'}x^{2}+2\alpha x)=0\ .
\]
From these three expressions, we know that the sum of magnetic pressure
and tension forces actually does not affect the dynamic behaviour at
very large $x$ (i.e., a nearly force-free condition). When $A>2$, the 
inward gravitational force overpowers the pressure gradient force. 
When $A=2$, the gravitational force is barely balanced by the pressure 
gradient force. The latter situation is closely pertinent to the EWCS 
of \citet{shu77}.

In the limit of $x\rightarrow 0^{+}$, the asymptotic
solutions for the central free-fall collapse is
\begin{equation}\label{class1v}
v\rightarrow -2F/x^{1/2}-\frac{3}{4F}x^{1/2}\ln x-2L
x^{1/2}+\cdots \ ,
\end{equation}
\begin{equation}\label{class1a}
\alpha\rightarrow F/x^{3/2}-\frac{3}{8F}x^{-1/2}\ln x-L
x^{-1/2}+\cdots \ ,
\end{equation}
\begin{equation}
\beta=\lambda\alpha^{2} x^{2}=\frac{\lambda F^2}{x}+\cdots \ ,
\end{equation}
where $F>0$ and $L$ are two constant parameters. We may take
$F=m_{0}/\sqrt{2}$ with parameter $m_{0}$ being a measure 
for the central mass accretion rate (Lou \& Shen 2004).
In dimensional form, the transverse magnetic energy density is
$$
{B_{\perp}^2\over 8\pi }\rightarrow 
{\lambda a^3m_{0}^2\over 16\pi Gtr}\ .
$$
%
The pressure gradient force in
the limit of $x\rightarrow0^{+}$ is
\[
\qquad -\frac{\alpha^{'}}{\alpha}=\frac{3}{2x}\ ,
\]
the gravitational force in the
limit of $x\rightarrow0^{+}$ is
\[
\qquad -\alpha(x-v)=-\frac{2F^{2}}{x^{2}}\ ,
\]
and the sum of the magnetic pressure and tension forces 
together in the limit of $x\rightarrow0^{+}$ is
\[
\qquad -\lambda(\alpha^{'}x^{2}+2\alpha x)=-\frac{\lambda F}{2x^{1/2}}\ .
\]
In the limit of $x\rightarrow0^{+}$, the most dominant force is the 
gravitational force leading to a gravitational core collapse. The 
pressure gradient force is the second dominant one. In comparison, 
the sum of the magnetic pressure and tension forces is inward and 
has the weakest effect as $x\rightarrow0^{+}$.






The other asymptotic solution for a core
expansion as $x\rightarrow 0^{+}$ is given by
\begin{equation}\label{class2v}
v\rightarrow
\frac{2}{3}x+\frac{(2-3D-18D\lambda)}{135}x^{3}+\cdots\ ,
\end{equation}
\begin{equation}\label{class2a}
\alpha\rightarrow D+\frac{D(2-3D-18D\lambda)}{18}x^{2}+\cdots\ ,
\end{equation}
\begin{equation}
\beta=\lambda\alpha^2 x^2=\lambda D^2 x^2+\cdots\ ,
\end{equation}
which is a fairly straightforward MHD generalization
of the Larson-Penston type solutions.

The pressure gradient force in
the limit of $x\rightarrow 0^{+}$ is
\[
-\frac{\alpha^{'}}{\alpha}=-\frac{(2-3D-18D\lambda)}{9}x\ ,
\]
the gravitational force in the
limit of $x\rightarrow 0^{+}$ is
\[
-\alpha(x-v)=-\frac{D}{3}x\ ,
\]
and the sum of the magnetic pressure and tension
forces in the limit of $x\rightarrow0^{+}$ is
\[
-\lambda(\alpha^{'}x^{2}+2\alpha x)=-2\lambda D x\ .
\]
By comparing the three forces, it is clear that the magnetic 
field plays an important dynamical role in the core expansion. 
This involves a radially inward net magnetic force. By setting 
$\lambda$ equal to zero, we readily recover the results of 
\citet{shu02}.
%
From equation (\ref{class2v}), we know that $v>0$, $v^{'}>0$ 
and $(v-x)<0$. The total time derivative of $u\equiv av(x)$ 
is nevertheless negative and 
and the magnetic energy density is simply
$$
\frac{B_{\perp}^{2}}{8\pi}=\frac{\lambda D^{2}r^{2}}{8\pi G t^{4}}\ .
$$
Please note that the straight line $x-v=0$ is another special 
demarcation line of importance, especially for the physical 
requirement of a positive $m(x)$. In the presence of magnetic field, 
the magnetosonic critical lines become curved in a characteristic 
manner. In contrast, the sonic critical line in the absence of 
magnetic field is a straight line $x-v=1$ in the first and fourth 
quadrants. When the magnetic field is included, the magnetosonic 
critical lines become curves in the first and fourth quadrants. 
In order to go across the magnetosonic critical line smoothly, the 
primary and secondary directions of the eigensolutions need to be 
reanalyzed accordingly. The topological characteristics of the 
saddle and nodal points have also been changed by the inclusion of 
magnetic field (see Appendix A for details).

\section{Numerical Solutions of
\\ \quad the Nonlinear MHD ODEs}

We derive global solutions in the semi-complete solution space from 
$x\rightarrow+\infty$ to $x\rightarrow0^{+}$. In fact, there exists 
a one-to-one correspondence between the semi-complete and complete 
solutions,
because of the invariance property for the time reversal symmetry
discussed earlier (see figure 6 of Lou \& Shen 2004 for examples). 
We shall solve below the coupled nonlinear MHD ODEs for similarity 
solutions making use of the magnetosonic critical line analyses and 
the asymptotic solution properties presented in \S\ 3. Numerically, 
we can construct various types of solutions that may or may not 
cross the magnetosonic critical line. Integrating from 
$x\rightarrow +\infty$ with three constant parameters $\lambda$, 
$V$ and $A$ specified in asymptotic solutions (\ref{infinity11}) 
and (\ref{infinity22}), we readily obtain those similarity solutions
without crossing the magnetosonic critical line.

For global smooth self-similar solutions satisfying critical
conditions $(\ref{critical1})-(\ref{critical3})$ along the
magnetosonic line, we need to determine the first derivatives
of $v$ and $\alpha$ using the L'H\^{o}spital rule. We then
integrate leftward (backward) and rightward (forward) from
the magnetosonic critical point. Based on the root properties
of quadratic equation (\ref{criticalderivative}) for eigenvalues
of the radial speed gradient $v'$, we can figure out the nature
of the magnetosonic critical points (e.g., Jordan \& Smith 1977;
Whitworth \& Summers 1985).
The magnetosonic critical line with $x-v>0$ covers the first and
fourth quadrants. Points along the magnetosonic critical line in the 
first quadrant [$0<x<(1+2\lambda)^{1/2}$] are saddle points, while 
points along the magnetosonic critical line in the fourth quadrant 
[$(1+2\lambda)^{1/2}<x<+\infty$] are nodal points (see Appendix A 
for details). In the first quadrant, type 1 eigensolutions correspond 
to negative $v$ gradients and type 2 eigensolutions correspond to 
positive first derivatives of $v$. In the fourth quadrant, type 1 
eigensolutions correspond to lower absolute values of first 
derivatives of $v$ and type 2 eigensolutions have larger absolute 
values of first derivatives of $v$ with both $v'$ being positive 
(see Appendix A). Numerical integrations away from a saddle point 
tend to be stable. Unstable or neutrally stable integrations may 
occur when integrating away from a nodal point 
\citep{whitworth85,hunter86,jordan77}. In the following we explore 
numerically various types of solutions that may or may not intersect 
the magnetosonic critical line.

\subsection[]{Global solutions without crossing the
\\ \ \qquad
Magnetosonic Critical Line}


The MHD generalization of EWCS of \citet{shu77} can be obtained
by joining two solution segments. One portion for a central MHD
free-fall collapse is to the left of $x=(1+2\lambda)^{1/2}$ and
the other for a magnetostatic quasi-spherical envelope is to the
right of $x=(1+2\lambda)^{1/2}$. The inner part of MHD EWCS
can be obtained by integrating leftward from $x=(1+2\lambda)^{1/2}$
with the first derivatives of $v$ and $\alpha$ given by $v_{c1}^{'}$
and $\alpha_{c1}^{'}$ in equations (\ref{ewcsv}) and (\ref{ewcsa}),
respectively. Note that first derivatives $v_{c2}^{'}$ and
$\alpha_{c2}^{'}$ cannot be used to integrate leftward. The right
portion of the MHD EWCS is the special analytical solution
(\ref{specialone}). In Figure 3, the MHD EWCS solution is plotted
in a heavy solid line. This solution can also be obtained by
taking the limit of $A\rightarrow2^{+}$. Note that the MHD EWCS
is tangential to the magnetosonic critical line at the point
$x=(1+2\lambda)^{1/2}$. At this point, $\alpha=2x^{-2}$ and thus
\[
\qquad (a^{2}+v_{A}^{2})/a^2=1+\lambda\alpha x^2=1+2\lambda\ .
\]
In other words, we have $
r/(at)=(a^{2}+v_{A}^{2})^{1/2}/a
$
corresponding to an MHD collapse wave front travelling
outward at the magnetosonic speed $(a^{2}+v_{A}^{2})^{1/2}$
(i.e., the fast MHD wave speed perpendicular to the local
magnetic field).



In star forming regions, e.g., the B335 cloud system, a 
typical mass density in molecular clouds is of the order of
$10^{-18}\hbox{ g/cm}^{3}$ and a typical magnetic field strength 
is around $100\mu$G. The Alfv\'en speed $v_{A}$ is approximately
$0.4\hbox{ km s}^{-1}$, which is comparable to the typical sound 
speed $a$ in molecular cloud cores with temperature of a few tens 
of Kelvin degrees. Therefore the inclusion of a tangled magnetic 
field would increase the central mass accretion rate by a factor 
of $\sim 3$ or so in the scenario of a magnetized EWCS as compared 
to the purely hydrodynamic EWCS. The overall dynamical timescale 
of the infalling envelope would be shortened to a few times 
$10^{4}$ yr.

\begin{table*}
\centerline{Variations of Accretion Rate with $\lambda$ }
\begin{tabular}{|c|c|c|c|c|}
  \hline
  \hline
  $\lambda$ & 0 & 0.1 & 1.0 & 2.0 \\
  \hline
  $m_{0}$ & 0.98 & 1.04 & 1.42 & 1.73\\
  \hline
\end{tabular}
\caption{Central mass accretion rates of MHD expansion-wave
collapse solution (marked by the heavy solid line in Fig. 3)
matched with an isothermal magnetostatic envelope. }
\end{table*}

The reduced mass density $\alpha(x)$ is continuous at the 
`kink' point for the magnetosonic collapse front. In the case 
of an MHD generalization of Shu (1977), the central accretion 
rate changes with the variation of $\lambda$. As $\lambda$ 
takes higher values, the core accretion rate becomes greater.

As $x$ approaches infinity, solutions should gradually 
converge to the asymptotic solutions (\ref{infinity11}) and
(\ref{infinity22}), while as $x$ approaches the origin, global 
solutions should satisfy solutions (\ref{class1v}), (\ref{class1a}),
(\ref{class2v}) and (\ref{class2a}). Those solutions without
crossing the magnetosonic critical line can be readily obtained by
numerically integrating from the asymptotic solutions
(\ref{infinity11}) and (\ref{infinity22}) at a sufficiently large
$x$. For those solutions crossing the magnetosonic critical line
analytically, the information about the magnetosonic critical line
and the gradients of $v$ and $\alpha$ at the critical point
is crucial for constructing the solutions.
\begin{figure}
 \includegraphics[width=3in]{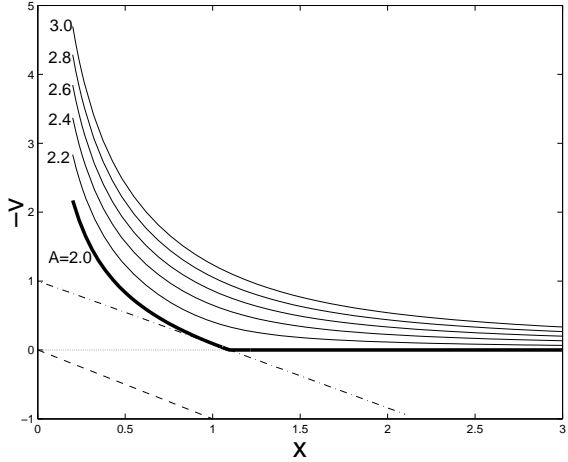}
 \caption{\label{v0without} As examples of illustration,
 we choose $\lambda=0.1$, $V=0$, $A=2.0,\ 2.2,\ 2.4,\ 2.6,\
 2.8,\ 3.0$ (from bottom to top in order along the solution 
 curves); the vertical coordinate is $-v$ and the horizontal 
 coordinate is $x$. This figure is qualitatively similar to 
 figure 1 of \citet{loushen}. With a magnetic field included, 
 the critical line has a different shape from that of the 
 pure hydrodynamic case. Dashed line is $x-v=0$. Dot-dashed 
 line is the curved magnetosonic critical line.  }
\end{figure}
\begin{figure}
 \includegraphics[width=3in]{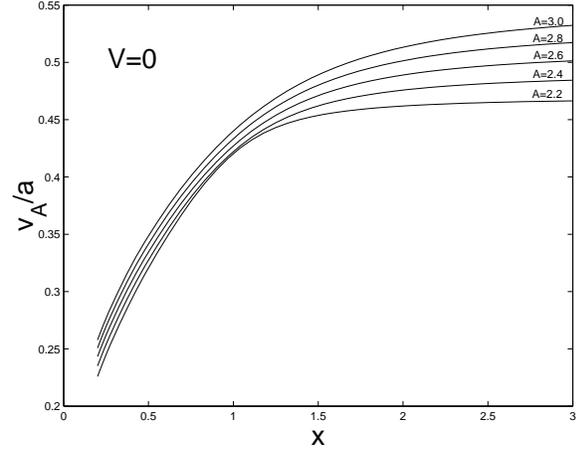}
 \caption{\label{v0alfven}We adopt $\lambda=0.1$ as in Figure 3.
 The corresponding Alfv\'en-to-sound speed ratios $v_{A}/a$ of
 similarity solutions with $V=0$, $A=2.0,\ 2.2,\ 2.4,\ 2.6,\
 2.8,\ 3.0$ (from bottom to top) are displayed; the vertical
 coordinate is the speed ratio $v_A/a$ and the horizontal 
 coordinate is $x$.  }
\end{figure}
\begin{figure}
 \includegraphics[width=3in]{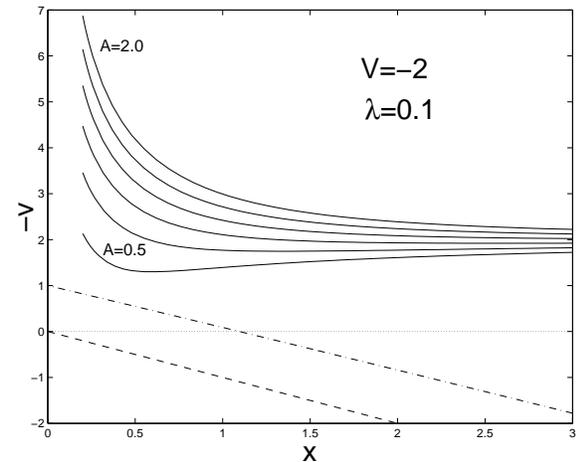}
 \caption{\label{vm2without}Examples of radial inflows at large
 $x$ with $\lambda=0.1$, $V=-2$, $A=0.5,\ 0.8,\ 1.1,\ 1.4,\ 1.7,\
 2.0$, respectively (from bottom to top); the vertical coordinate
 is $-v$ and the horizontal coordinate is $x$. The straight dashed 
 line is $x-v=0$. Dot-dashed line is the magnetosonic critical line. }
\end{figure}
\begin{figure}
 \includegraphics[width=3in]{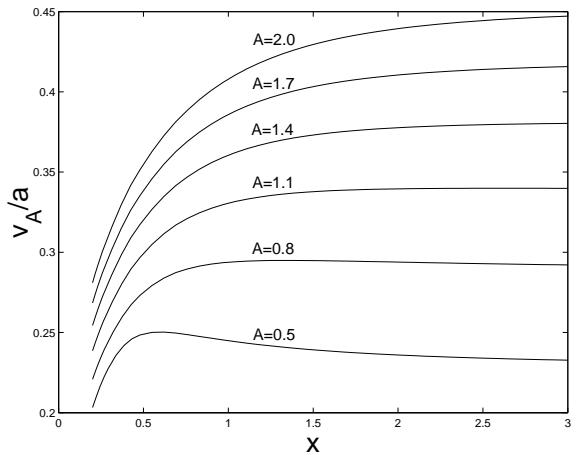}
 \caption{\label{vm2alfven}Here $\lambda=0.1$ as in Figure 5.
 The corresponding speed ratios of $v_{A}/a$ for solutions 
 with $V=-2$, $A=0.5,\ 0.8,\ 1.1,\ 1.4,\ 1.7,\ 2.0$ 
 respectively (from bottom to top); the vertical coordinate is 
 the speed ratio $v_A/a$ and the horizontal coordinate is $x$.  }
\end{figure}
\begin{figure}
 \includegraphics[width=3in]{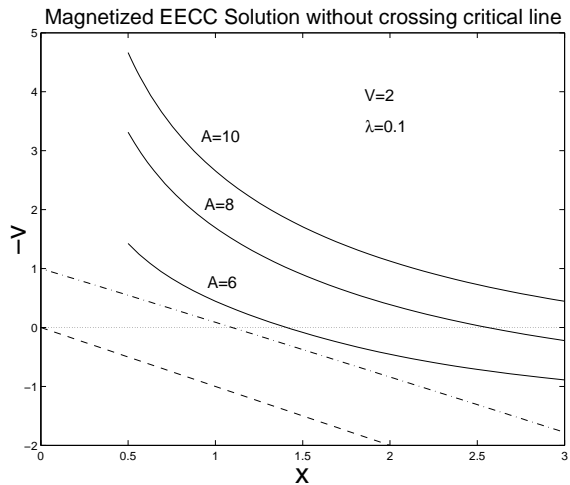}
 \caption{\label{v2without}For $\lambda=0.1$, $V=2$, and
 $A=6,\ 8,\ 10$, respectively (from bottom to top), we 
 readily obtain MHD EECC similarity solutions without 
 crossing the magnetosonic critical line; the vertical 
 coordinate is $-v$ and the horizontal coordinate is $x$.
 The stagnation point with $-v=0$ travels outward with a 
 super magnetosonic speed. The straight dashed line is 
 $x-v=0$. The dash-dotted curve is the magnetosonic 
 critical line. }
\end{figure}
\begin{figure}
 \includegraphics[width=3in]{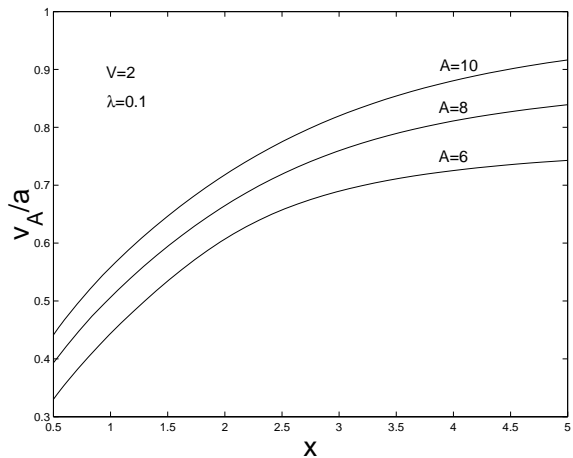}
 \caption{\label{v2alfven}For $\lambda=0.1$ as in Figure 7,
 we display the corresponding speed ratios $v_A/a$ for
 solutions with $V=2$ and $A=6,\ 8,\ 10$, respectively (from
 bottom to top); the vertical coordinate is the speed ratio
 $v_A/a$ and the horizontal coordinate is $x$.
  }
\end{figure}

In Figures $3-5$, we integrate the coupled nonlinear ODEs
(\ref{mainv}) and (\ref{mainalpha}) from a large enough $x$ with 
$V$, $A$ and $\lambda$ specified and show global solutions without 
crossing the magnetosonic critical line. In Figure 3, we numerically 
integrate the equations with $V=0$ (the so-called asymptotic breeze 
solutions; Shen \& Lou 2004).
For $V<0$ and $V>0$ corresponding to inflows
and outflows at $x\rightarrow+\infty$ respectively,
self-similar solutions can be readily constructed.
For a specified value of speed parameter $V$, the mass parameter $A$
should meet certain criteria in order to have global solutions. Due
to the nonlinearity of these ODEs and in the presence of magnetic
field, this problem is fairly complex and solutions are explored
through trial and error \citep{loushen}.

We now summarize some results of our numerical exploration. For 
numerical solutions with $V=0$, condition $A>2$ is necessary for 
self-similar solutions to exist without crossing the magnetosonic 
critical line in the first quadrant. For $V<0$, the lower limit 
of $A$ for possible similarity solutions without crossing the 
magnetosonic critical line is less than 2. In our illustrating 
examples with $\lambda=0.1$, the lower limit of $A$ is $\sim 0.39$. 
For $A\lsim 0.39$, similarity MHD solutions would crash onto the 
magnetocritical line. For $V>0$ and $\lambda=0.1$, the minimum 
value of $A$ for similarity MHD solutions without crossing the 
magnetosonic critical line now becomes generally greater than 
$\sim 2$, and varies for different values of $V$. In our specific 
computations with $\lambda=0.1$ and $V=2$, the minimum value of 
$A$ is $\sim 5.6$. For $A\lsim 5.6$, the MHD solutions would crash 
onto the magnetocritical line.

\subsection[]{Analytical Solutions for Crossing
\\ \qquad the Magnetosonic Critical Line}

\subsubsection[]{Type 1 Similarity Solutions}

We first describe the type 1 similarity solutions. By integrating
leftward and rightward from a point along the magnetosonic critical
line with a negative first derivative of $v(x)$, we obtain the type 1
self-similar solutions. These solutions of $v(x)$ in the first quadrant,
referred to as the plus solutions by \citet{shu77}, approach negative
constant values of $V$ as $x\rightarrow+\infty$. Following the same
procedure, we can also construct type 1 solutions which pass through 
the magnetosonic critical line in the fourth quadrant. These solutions 
have the properties that $v(x)$ approaches a positive constant $V$ at 
large $x$. We shall see that a class of discrete yet infinitely many 
type 1 solutions matches finite solutions (\ref{class2v}) and 
(\ref{class2a}) near the origin $x=0$. This type of solutions were 
first derived by \citet{hunter77}. Our analysis generalizes these 
self-similar solutions by including a tangled magnetic field. For 
further generalizations of LP solutions with various shocks, the
interested reader is referred to the recent work of Bian \& Lou (2005).

The plus solutions of Shu (1977) (or type 1 solutions, see
solution curves in Figure 6) with positive $d(-v)/dx$ in the first
quadrant have their counterparts in the presence of magnetic field.
They pass through the magnetosonic critical line and approach
a constant negative speed $V$ at large $x$. These solutions may
be matched with an approximate quasi-magnetohydrostatic state.
As $x\rightarrow 0^{+}$ and $v\rightarrow 0$,
we derive the following equation governing
the quasi-magnetohydrostatic state, namely
\[
\frac{1}{x^2}\frac{d}{dx}
\bigg\lbrace x^2\bigg[-(1+\lambda\alpha x^2)
\bigg(\frac{\alpha^{'}}{\alpha}\bigg)
-2\lambda\alpha x\bigg]\bigg\rbrace=\alpha\ .
\]
For $\lambda=0$, this equation reduces to the Lane-Emden equation
(see relevant discussions in the Appendix of Shu 1977).
These solutions represent self-similar inflows with a constant speed
at large $x$. Besides matching with quasi-magnetostatic solutions,
it is possible to match these solutions with self-similar outflows
at smaller $x$ with or without shocks (Tsai \& Hsu 1995; Shu et al.
2002; Shen \& Lou 2004; Bian \& Lou 2005).
\begin{figure}
 \includegraphics[width=3in]{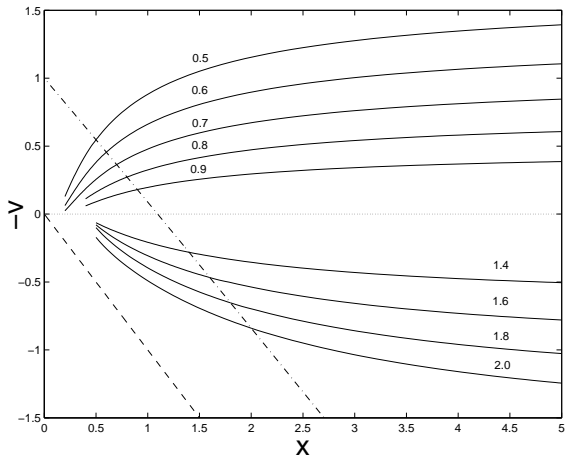}
 \caption{\label{type1cross}For $\lambda=0.1$, we show two 
 classes of examples for type 1 solutions smoothly crossing 
 the magnetosonic line in the first and fourth quadrants, 
 separately. The critical points are $x_{*}=0.5,\ 0.6,\ 
 0.7,\ 0.8,\ 0.9,\ 1.4,\ 1.6,\ 1.8,\ 2.0$, respectively. 
 The straight dashed line is $x-v=0$ and the dash-dotted 
 curve is the magnetosonic critical line.  }
\end{figure}
\begin{figure}
\includegraphics[width=3in]{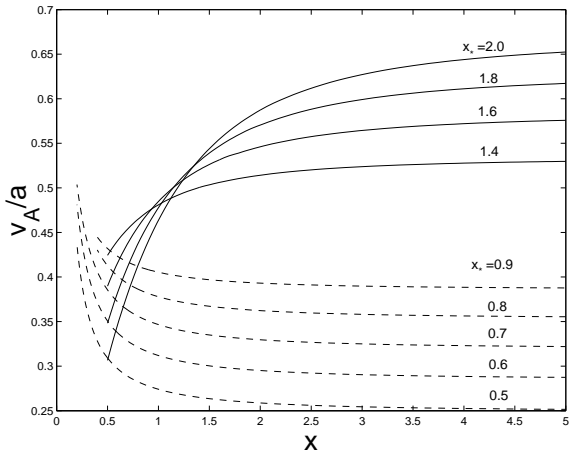}
\caption{\label{type1alfven1} For $\lambda=0.1$ as in Figure 9,
the corresponding Alfv\'en to sound speed ratios $v_{A}/a$ of
type 1 solutions that smoothly cross the magnetosonic critical 
line in the first quadrant. The chosen critical points are 
$x_{*}=0.5,\ 0.6,\ 0.7,\ 0.8,\ 0.9,\ 1.4,\ 1.6,\ 1.8,\ 2.0$.  }
\end{figure}

\subsubsection[]{Self-Similar MHD Solutions crossing
the Magnetosonic Critical Line Analytically }

To find the solutions crossing the magnetosonic critical line
analytically, we follow the same procedure of Hunter (1977)
and Lou \& Shen (2004).
Choosing a meeting point $x_{m}$ and from different first critical
point $x_{*}(1)$ [$x_{*}(1)<(1+2\lambda)^{1/2}$ and $x_m>x_{*}(1)$],
we integrate the coupled nonlinear MHD ODEs forward towards the
meeting point $x_{m}$ using the type 2 eigensolution. With every 
$x_{*}(1)$, we obtain a pair $\{ v,\alpha\}$ at the meeting point 
$x_{m}$ in the so-called phase diagram of $v$ versus $\alpha$. For 
a series of $x_{*}(1)$, we obtain a series of $\{v,\alpha\}$ pairs 
and thus a curve in the phase diagram.
Similarly, we can integrate the coupled nonlinear MHD ODEs from
different $x_{*}(2)$ [$x_{*}(2)>(1+2\lambda)^{1/2}$ and
$x_{*}(2)>x_m$] backward towards the same meeting point $x_{m}$ to
obtain another curve in the $\{v,\alpha\}$ phase diagram. When we
integrate from $x_{*}(2)$ towards the meeting point $x_m$, both
type 1 and type 2 solutions may be chosen to construct such
solutions, while integration from $x_{*}(1)$ necessarily requires
the type 2 solution. 
Regarding the stability of numerical integrations, we note that to 
go from $x_{*}(2)$ towards the meeting point $x_m$ using type 1 
solutions, the integration is away from a node along the secondary 
direction and is neutrally stable. As we go from $x_{*}(2)$ towards 
the meeting point $x_m$ using type 2 solutions, the integration is 
away from a node along the primary direction and is unstable and 
extremely sensitive to the accuracy of the starting conditions. 
Based on intersections of the two curves in the $\alpha-v$ phase 
diagram, we identify matches for global similarity solutions. From 
these intersection points, we determine the relevant $x_{*}(1)$ 
and $x_{*}(2)$ and compute solutions. Global similarity solutions 
obtained this way cross the magnetosonic critical line twice 
analytically.

In Figure \ref{phasediagram}, we present such a phase diagram
of $v$ versus $\alpha$ following the procedure described above.
The first three intersection points for $v>-0.5$ in the phase
diagram lead to the first three global solutions that cross
the magnetosonic critical line twice analytically. For the
first solution, we have $0.10306<x_{*}(1)<0.10307$ and
$1.81092<x_{*}(2)<1.81093$. For the second solution, we have
$0.000128<x_{*}(1)<0.000129$ and $0.8729<x_{*}(2)<0.8730$.
For the third solution, we have $1.2\times10^{-6}<x_{*}(1)
< 1.3\times10^{-6}$ and $1.165<x_{*}(2)<1.166$.

In Figure \ref{type2type1-123}, we display the first three global
solutions that go across the magnetosonic critical line twice
analytically. Similar oscillation phenomena occur in magnetized
cases as those in purely hydrodynamic cases \citep{loushen}. In
Figure \ref{type2type1blowup}, we present the enlarged part in
Figure \ref{type2type1-123} for small $x$ with a logarithmic
scale.

\begin{figure}
 \includegraphics[width=3in]{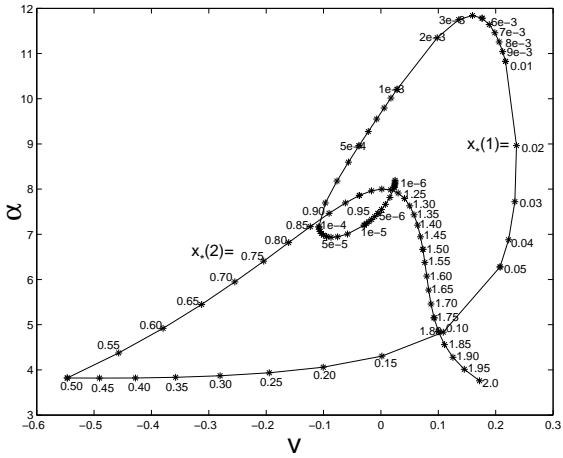}
 \caption{\label{phasediagram}The case of $\lambda=0.1$. 
 The type 2-type 1
 solutions matching phase diagram of $v$ versus $\alpha$. The 
 meeting point is chosen at $x_{m}=0.5$. The type 2 eigenvalue 
 is either positive in the $v$ gradient (saddle point) or of 
 larger absolute value for the $v$ gradient (nodal point). The 
 type 1 is negative in the $v$ gradient (saddle point) or of 
 smaller absolute value for the $v$ gradient (nodal point). 
 The $x_{*}(1)$ and $x_{*}(2)$ of the first solution of this 
 type are in the range $0.10306<x_{*}(1)<0.10307$ and 
 $1.81092<x_{*}(2)<1.81093$. For the second solution of this 
 type, we have $0.000128<x_{*}(1)<0.000129$ and $0.8729<x_{*}(2)
 <0.8730$. As for the third solution of this type, we have
 $1.2\times10^{-6}<x_{*}(1)<1.3\times10^{-6}$ and $1.165<x_{*}(2)<1.166$.
}
\end{figure}

\begin{figure}
 \includegraphics[width=3in]{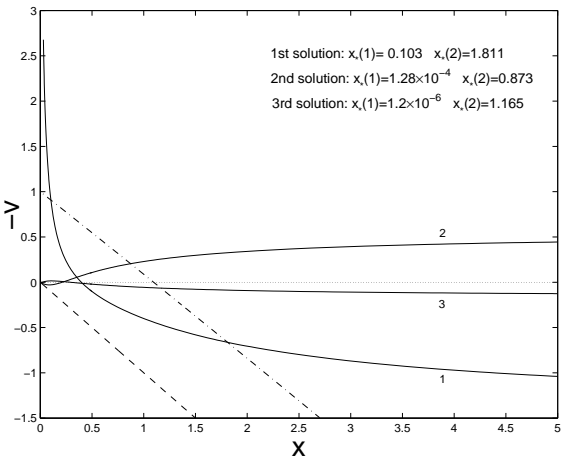}
 \caption{\label{type2type1-123} The case of $\lambda=0.1$. The
 first three solutions passing across the magnetosonic critical 
 line twice with increasing number of stagnation points moving 
 outward with sub-magnetosonic speeds. We have $x_{*}(1)\approx 
 0.10306$ and $x_{*}(2)\approx 1.81092$ for the first solution 
 with one stagnation point; $x_{*}(1)\approx 1.28\times 10^{-4}$ 
 and $x_{*}(2)\approx 0.873$ for the second solution with two 
 stagnation points; and $x_{*}(1)\approx 1.2\times 10^{-6}$ 
 and $x_{*}(2)\approx 1.165$ for the third solution with three 
 stagnation points. The straight dashed line is $x-v=0$ and 
 the dash-dotted curve is the magnetosonic critical line.   }
\end{figure}
%

\begin{figure}
 \includegraphics[width=3in]{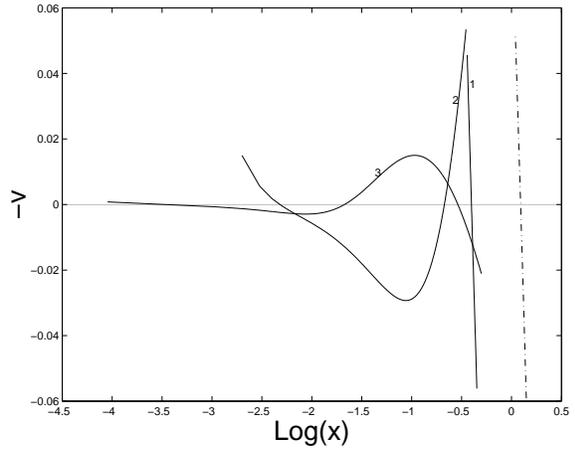}
 \caption{\label{type2type1blowup} The case of $\lambda=0.1$. The 
 vertical coordinate is $-v$ and the horizontal coordinate is 
 $\log(x)$. By this enlargement, we see clearly that first the three 
 solutions have one, two, and three stagnation points, respectively. 
 The dash-dotted line is the magnetosonic critical line.  }
\end{figure}

\begin{figure}
 \includegraphics[width=3in]{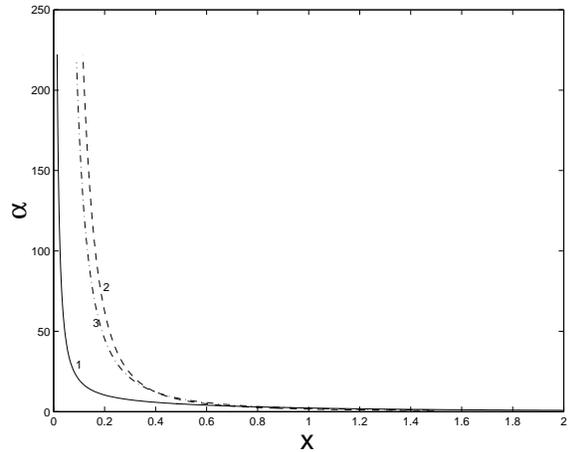}
 \caption{\label{type2type1alpha123} The case of $\lambda=0.1$. The 
 vertical coordinate is $\alpha$ and the horizontal coordinate is 
 $x$. The first three solutions go across the magnetosonic critical 
 line analytically. Solid, dashed and dash-dotted curves correspond 
 to the first, second, and third solutions, respectively.  }
\end{figure}

\subsubsection[]{Larson-Penston (LP) Type
Solutions and Hunter Type Solutions}

\citet{larson69a} and \citet{penston69a} independently derived
a kind of monotonic similarity solution that has the asymptotic
behaviour (\ref{class2v}) and (\ref{class2a}) with $\lambda=0$
(i.e.,  without magnetic field) near the origin $x\rightarrow 0^{+}$.
In the presence of magnetic field, the MHD extension of LP solution
has the asymptotic behaviour (\ref{class2v}) and (\ref{class2a})
with $\lambda>0$. The MHD LP-type solution can be obtained similarly
by trying different values of $D$ with a given $\lambda$ near the
origin $x\rightarrow 0^{+}$ and by integrating nonlinear MHD ODEs
(\ref{mainv}) and (\ref{mainalpha}) rightward from $x=0$ towards
the magnetosonic critical line $(x-v)+2\lambda x=(x-v)^{3}$ until
the critical conditions (\ref{critical1})$-$(\ref{critical3}) are
satisfied in such a way that the solution matches with one of the
two eigensolutions across the magnetosonic critical line. Thus
the problem of finding global LP-type solutions is basically an
eigenvalue problem of determining discrete eigenvalues for the 
$D$ parameter.

For the hydrodynamic LP solution, the eigenvalue $D$ is
approximately 1.67. In our modified definition of type 1 and
type 2 solutions in the presence of magnetic field, the
corresponding eigensolutions are related to type 1 solutions
[n.b., not type 2 solutions, see \citet{shu77,hunter77,
whitworth85,loushen}].\footnote{In previous studies \citep{shu77,
hunter77,whitworth85,loushen}, while the LP solution is formally
classified as type 2 solution (cf. our terminology) across the
sonic critical line in the fourth quadrant, the corresponding
eigensolution is actually along the secondary direction for
$x_{*}>2$. Our type 1 and type 2 definitions would not suffer
from the disadvantage of magnitude reversal in the sense of
$dv/dx$. By our convention, type 1 is always along the secondary
direction, and type 2 is always along the primary direction.}
\begin{figure}
 \includegraphics[width=3in]{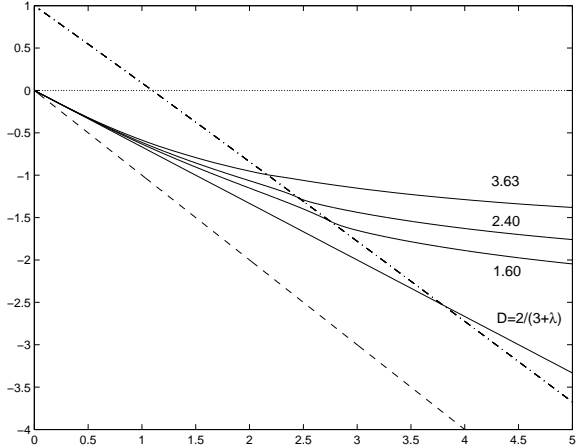}
 \caption{\label{lpsolution} The case of $\lambda=0.1$.
 $D=2/(3+18\lambda)$, 1.60, 2.40, 3.63 running from lower 
 to upper solution curves. LP-type solution passes across 
 the magnetosonic critical line. The straight dashed line 
 is $x-v=0$. The slightly curved dash-dotted line is the 
 magnetosonic critical line.
  }
\end{figure}
LP-type solution goes across the magnetosonic critical line
analytically. From our numerical exploration for $\lambda=0.1$, the
corresponding eigenvalue is $D\cong 3.63$. If we relax the constraints
that solutions must be analytic (as defined by Hunter 1986) across the
magnetosonic critical line, it is then possible to construct continuum
bands of solutions crossing the magnetosonic critical line with weak
discontinuities \citep{whitworth85}. The case of $D=2/(3+18\lambda)$
corresponds to the special exact global solution (\ref{specialtwo}).


Hunter type solutions refer to those similarity solutions crossing 
the magnetosonic critical line analytically and having asymptotic 
behaviour (\ref{class2v}) and (\ref{class2a}) as $x$ approaches $0$. 
These solutions pass through the magnetosonic critical line along 
the secondary direction, i.e., they pass the magnetosonic critical 
line as type 1 solutions. We choose a meeting point at $x_{m}=0.5$. 
From around the origin $x\rightarrow 0^{+}$ with $D$ and $\lambda$ 
specified in equations (\ref{class2v}) and (\ref{class2a}), we 
integrate the nonlinear MHD ODEs to the meeting point $x_{m}=0.5$. 
For every $D$, we obtain a $\{v,\alpha\}$ pair at $x_{m}$ in the 
phase diagram. By varying $D$ parameter in sequence, we readily
determine a series of $\{v,\alpha\}$ pairs. Plotting this series
of $\{v,\alpha\}$, we get a curve in the phase diagram. Meanwhile,
we integrate the nonlinear MHD ODEs from different critical points
$x_{*}$ (with $x_{*}>x_{m}$) along the magnetosonic critical line
backward towards the same meeting point $x_{m}$ to get another
curve in the $\{\alpha,\ v\}$ phase diagram. Once the intersection
points are determined in this phase diagram, we readily obtain the
corresponding Hunter type solutions.

\begin{figure}
 \includegraphics[width=3in]{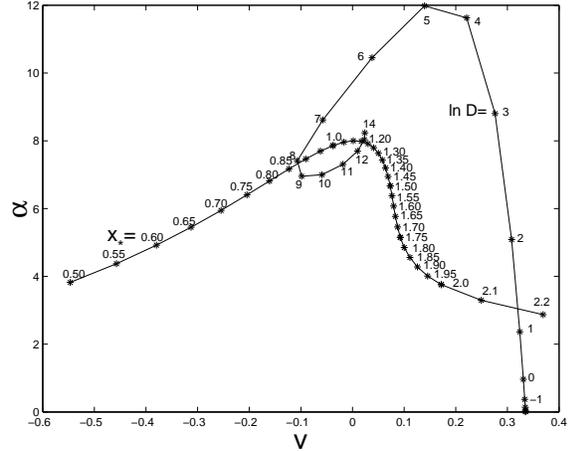}
 \caption{\label{hunterphasedia}The case of $\lambda=0.1$. The
 meeting point is chosen at $x_{m}=0.5$. The two parameters $D$
 and $x_{*}$ for the first solution of this type are in the ranges
 of $1.29<\ln D<1.30$ and $2.164<x_{*}<2.165$, respectively. For
 the second solution of this type, we have $8.15<\ln D<8.16$ and
 $0.8704<x_{*}<0.8705$, respectively. And for the third solution
 of this type, we have $12.77<\ln D<12.78$ and $1.165<x_{*}<1.166$,
 respectively.  }
\end{figure}

\begin{figure}
 \includegraphics[width=3in]{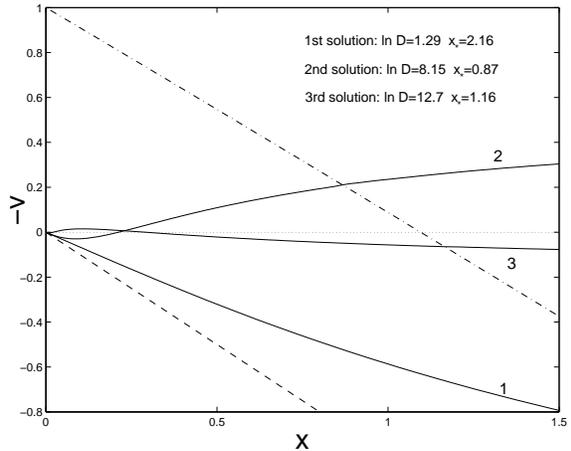}
 \caption{\label{hunterglobal}
 The case of $\lambda=0.1$ corresponding Fig. \ref{hunterphasedia}. 
 The first three Hunter type global solutions acrossing the 
 magnetosonic critical line. The straight dashed line is $x-v=0$. 
 The dash-dotted curve is the magnetosonic critical line. For the 
 first solution, we have $\ln D=1.29$ and $x_{\ast}=2.16$; for the 
 second solution, we have $\ln D=8.15$ and $x_{\ast}=0.87$; and for 
 the third solution, we have $\ln D=12.7$ and $x_{\ast}=1.16$.  }
\end{figure}


\section{Notes and Discussion}

In this paper, we investigated quasi-spherical self-similar
flow solutions for magnetized self-gravitating isothermal fluids
in the semi-complete solution space. The relevant MHD similarity
solutions are obtained and classified by comparing with results
of previous analyses \citep{larson69a,penston69a,shu77,hunter77,
whitworth85,chiueh,loushen,shenlou04}. The straight sonic critical 
line is now replaced by the curved magnetosonic critical line as 
expected physically. Global similarity solutions without crossing the
magnetosonic critical line also show ECCC and EECC characteristics,
and are fairly easy to construct numerically in the isothermal
approximation. In a polytropic gas with radiative loss effects,
some qualitatively similar features were briefly mentioned in
\citet{boily95}. It is also possible to construct similarity
solutions for a magnetized polytropic gas with a quasi-spherical
symmetry on larger scales (Wang \& Lou 2005, in preparation).

With magnetic field included, both magnetized expansion wave collapse
solutions (mEWCS) and magnetized envelope expansion core collapse
(mEECC) solutions can be readily constructed. The mEWCS structure is
composed of two main parts joined at the collapse wave front $x_{*}=
(1+2\lambda)^{1/2}$. The outer part with $x\geqslant(1+2\lambda)^{1/2}$
is a nearly force-free magnetostatic envelope with the magnetic pressure
and tension forces almost cancelling each other and the inner part with
$x<(1+2\lambda)^{1/2}$ is an MHD core collapse characterized by diverging
behaviours as $x\rightarrow 0^{+}$. On the whole, such a mEWCS physically
describes a situation with an MHD core collapse approaching a free-fall
state in the very central region and with a magnetostatic exterior at
rest for $x\geqslant(1+2\lambda)^{1/2}$; the quasi-spherical
stagnation surface separating the interior and exterior expands
at the magnetosonic speed $(v_{A}^2+a^2)^{1/2}$.

For more general mEECC similarity solutions with only one stagnation
point $x^{*}$ along the $x-$axis, the interior collapse ($x<x^{*}$)
and the exterior expansion ($x\geqslant x^{*}$) concur in a
self-similar manner.
The outward expansion of the exterior envelope gives rise to a
wind of constant speed at large radii. The quasi-spherical surface
of separation where the gas remains at rest travels outward as
time goes on, similar to the `expansion-wave front' of Shu (1977).
From a more general perspective, the EWCS of Shu (1977) is a
special case of our mEECC similarity solutions in the sense that
an unmagnetized outer envelope is static at large $x$ instead of
expanding or contracting at constant speeds. The travel speed of
the stagnation front at $x^{*}$ could be either sub-magnetosonic
for $x^{*}<(1+2\lambda)^{1/2}$ or super-magnetosonic for
$x^{*}\geqslant(1+2\lambda)^{1/2}$.

This explanation can be straightforward extended to those EECC
solutions with more than one stagnation points, involving
self-similar radial oscillations in sub-magnetosonic region.
For such mEECC similarity solutions, there exist more than one
quasi-spherical stagnation surfaces of separation that expand
outward with time at different constant speeds. The separations
between quasi-spherical surfaces increase with time in a
self-similar manner.


Such mEECC similarity solutions, which depict an outgoing envelope and
a collapsing core, are derived from nonlinear MHD equations and may be
applied to various astrophysical processes with proper adaptions.
For star forming regions in molecular clouds, for instance, the starless
cloud system B335, we may take an infalling region of spatial scale
$\sim 1.5\times10^{4}$ AU, a cloud mass of $\sim 4M_{\odot}$, a magnetic
field strength of $\sim 134\mu$G \citep{wolf03} and a thermal sound speed
$a\sim 0.23\hbox{ km s}^{-1}$. In this case, the parameter $\lambda$
is approximately 0.01. By applying the EECC solution that crosses the
magnetosonic critical line twice analytically with one stagnation point
to model B335 system, we have the following estimates. As cloud
materials collapse from a radial distance $7700$ AU to $2200$ AU in
our frozen-in scenario, the magnetic field strength at $2200$ AU
would become $\sim 300\mu$G and the inflow speed at $2200$AU is
$\sim 0.11\hbox{ km s}^{-1}$. This inward increase of `primordial
magnetic field' may serve as an important `seed magnetic field'
for a further enhancement through MHD dynamo processes within the
central core of a protostar; the eventual appearance of generic
bipolar outflows which are most likely magnetized may result
from rotation and symmetry breaking of gross magnetic field
configurations. For central collapse regions very much closer to
the protostar, our self-similar solutions are not expected to be
applicable as transients and feedbacks lead to complicated
situations.


Such mEECC similarity solutions with embedded magnetic fields may
also be applied to the asymptotic giant branch (AGB) phase or
post-AGB phase in the late evolution stage of a star before the
gradual emergence evolution of a planetary nebula system with 
a central white dwarf of a high temperature. The timescale of 
this evolution `gap' is estimated to be $\sim 10^3$yrs. By the
conventional wisdom, a main sequence star with a progenitor of
less than several solar masses swells tremendously in size and
sustains a massive wind with an asymptotic speed of $\sim$ $10-20$
$\hbox{km s}^{-1}$; the mass loss rates fall within the range of
$\sim 10^{-8}$ to $\sim 10^{-4}M_{\odot}\hbox{ yr}^{-1}$. With
an insufficient nuclear fuel supply from a certain epoch on, the
central region starts to collapse while the outer envelope continues
to expand; it is highly likely that the central collapse is
accompanied by an outward energetic surge to chase the slowly
expanding envelope. Simultaneously, the outer expansion removes
stellar envelope mass, while the central infall and collapse produce
a proto white dwarf at the center. Sufficiently far away from the
initial and boundary conditions, the system may gradually evolve
into a dynamic phase representable by an EECC similarity solution
during a timescale of a few hundred to several thousand years. Within
this rough range of timescale, the accumulation of central core mass
should not exceed the Chandrasekhar mass limit of $1.4M_{\odot}$ for
a remnant white dwarf to survive; otherwise, a central explosion
(e.g., novae or even a supernova) may occur within a planetary nebula.
As a result of the energetic surge, the outer envelope expansion during
the EECC phase may be faster than the pre-existing massive slow wind.
This process will generally give rise to the formation of outward
moving shock when a faster EECC outflow catches up with a slower
pre-existing wind. Theoretically, it is possible to construct EECC
similarity shock solutions (Shen \& Lou 2004; Bian \& Lou 2005) to
generalize the earlier models of Tsai \& Hsu (1995) and Shu et al.
(2002). We therefore hypothesize that dynamical evolution of an
mEECC phase of around or less than a few thousand years may be the
missing linkage between the AGB or post-AGB phase and the gradual
appearance of a planetary nebula.
Depending on physical parameters of the progenitor star, it might
well happen that the core collapse and subsequent central infalls
during an mEECC phase lead to an eventual core mass exceeding the
Chandrasekhar mass limit of $1.4M_{\odot}$ and thus induce an
explosion with an intensity determined by the actual rate of 
central mass accretion.

For gross estimates, observations indicate a magnetic field strength of
$\sim 1$G near the stellar surface at $r\sim 1$AU during the AGB phase,
decreasing to $\sim 10^{-3}$G at $r\sim 10^3$AU. In the self-similar
expansion regime of our model analysis, we have the transverse magnetic
field $B_{\perp}\propto r^{-1}$ (e.g., Lou 1993, 1994), consistent with
the measured  magnetic field variations observed for AGB stars.
Much stronger surface magnetic fields of a white dwarf might be
generated and sustained by intrinsic MHD dynamo processes caused
by convective differential rotation within the interior of an
AGB star (e.g., Blackman et al. 2001).

The transverse magnetic field strength of a mEECC solution 
in the innermost free-fall collapse region behaves as 
$B_{\perp}\propto r^{-1/2}$. For a magnetic field strength 
of $1$G at $r\sim 1$AU and the radius of a proto-white dwarf 
of $\sim 6000\ $km, the order of magnitude in spatial scale 
of the system decreases roughly by a factor of $\sim 10^{6}$ 
and accordingly the magnetic field strength increases roughly 
by a factor of $\sim 10^{3}$. As a result, the magnetic field 
estimated at the surface of a proto-white dwarf is roughly 
of the order of $\sim 10^{3}\ $G. 

For the Crab Nebula as an example of supernova remnant, we simply
model the magnetic field as $B_{\perp}\propto r^{-1}$ during the
envelope expansion phase, ignoring complex interactions between
the pulsar wind and the inner nebula involved in the expansion
of the nebula. For a neutron star with a radius of $\sim 10$ km 
and an initial dipolar magnetic field strength of $\sim 10^{10}\ $G, 
the magnetic field would decrease to $\sim 10^{-3}\ $G at about 
several parsecs. These estimates roughly agree with the envelope 
expansion portion of our simple mEECC solutions.

For astrophysical systems of much larger scales such as
clusters of galaxies (e.g., Sarazin 1988; Fabian 1994),
the similarity solutions may be valuable for modeling a 
certain phase of their dynamical evolution involving 
large-scale random magnetic fields. Specifically, if their 
evolution involves a self-similar phase of central collapse
\citep{gunn72,fillmore84,bertschinger85,navarro97}, then
isothermal similarity solutions (\ref{class1v}), (\ref{class1a}),
(\ref{class2v}) and (\ref{class2a}) seem to suggest two possible
classes of galaxy clusters that release X-ray emissions through 
hot electrons virialized in gravitational potential wells. The 
bifurcation of extremely high X-ray core luminosities and normal 
X-ray core luminosities may correspond to steeper and shallower 
gravitational potential wells, respectively. Typically, a galaxy 
cluster is characterized by a size of order of a few Mpc, a mass 
of order of $10^{13}M_{\odot}$, and a magnetic field strength of 
a few $\mu $G.

For a galaxy cluster resulting from a collapse of a gravitationally
bound system on much larger scales, a primodial weak magnetic field
frozen into the gas can be enhanced by gradual contractions. In the
central core collapse region, a transverse magnetic field may scale
as $B\propto r^{-1/2}$. In this scenario, specific values of
primordial magnetic fields would be determined by the spatial scale
when the system starts a self-similar collapse.

While our mEECC and mECCC models are highly simplified and idealized, 
it would presumably grasp a few fundamental aspects of certain 
physical processes and provide valuable physical insights into the 
magnetic field evolution in various astrophysical systems.

\section*{Acknowledgments}
This research has been supported in part by the ASCI Center
for Astrophysical Thermonuclear Flashes at the University of
Chicago under the Department of Energy contract B341495, by the
Special Funds for Major State Basic Science Research Projects of 
China, by the Tsinghua Center for Astrophysics, by the Collaborative 
Research Fund from the National Science Foundation of China (NSFC) 
for Young Outstanding Overseas Chinese Scholars (NSFC 10028306) at 
the National Astronomical Observatories, Chinese Academy of Sciences, 
by the NSFC grant 10373009 at the Tsinghua University, and by the 
Yangtze Endowment from the Ministry of Education at the Tsinghua 
University. The hospitalities of the Mullard Space Science 
Laboratory at University College London and of Centre de Physique
des Particules de Marseille (CPPM/CNRS) + Universit\'e de la 
M\'editerran\'ee are also gratefully acknowledged. Affiliated 
institutions of YQL share this contribution.

\appendix
\section[]{ Analysis of Saddle
Points and Nodal Points }

This appendix A is to show explicitly which portion of the magnetosonic
critical line corresponds to saddle points and which portion of the
magnetosonic critical line corresponds to nodal points (e.g., Jordan
\& Smith 1977). We examine equation (\ref{criticalderivative}) in order
to extract information about the signs of the roots of equation
(\ref{criticalderivative}) along the magnetosonic critical line. In the
hydrodynamical case, the roots of equation (\ref{criticalderivative})
can be expressed in fairly simple analytical forms. We know that a point
within the open interval $0<x<1$ along the magnetosonic critical line 
is a saddle point, while a point within $x>1$ along the magnetosonic 
critical line is a nodal point.

In the presence of magnetic field, saddle and nodal points 
along the magnetosonic critical line are not so obvious as 
for the cases without magnetic field. Here, we perform such 
an analysis. Following the procedure of \citet{whitworth85}, 
we cast the self-similar nonlinear MHD ODEs in the forms of
\begin{equation}
dv/dx=Y(x,v,\alpha)/X(x,v,\alpha)\ ,
\end{equation}
\begin{equation}
d\alpha/dx=Z(x,v,\alpha)/X(x,v,\alpha)\ ,
\end{equation}
where functionals $X$, $Y$, and $Z$ are defined by
\begin{equation}
X(x,v,\alpha)\equiv (x-v)^2-(1+\lambda\alpha x^{2})\ ,
\end{equation}
\begin{equation}
Y(x,v,\alpha)\equiv\alpha(x-v)^2-2(x-v)/x\ ,
\end{equation}
\begin{equation}
Z(x,v,\alpha)\equiv\alpha^{2}(x-v)+2\lambda
x\alpha^{2}-2\alpha(x-v)^{2}/x\ .
\end{equation}
We figure out local topological properties of the solutions
by examining the eigenvalues of the following matrix
\[
M=\left(%
\begin{array}{ccc}
 \partial X/\partial x&\partial X/\partial v&\partial X/\partial\alpha\\
 \partial Y/\partial x&\partial Y/\partial v&\partial Y/\partial\alpha\\
 \partial Z/\partial x&\partial Z/\partial v&\partial Z/\partial\alpha\\
\end{array}%
\right)
\]
with the partial derivatives of $X$, $Y$ and $Z$ evaluated at 
a given point along the magnetosonic critical line. Detailed 
calculations show that one of the three eigenvalues is zero
and the other two eigenvalues satisfy the following quadratic 
equation
\begin{equation}\label{quadratic1}
az^{2}+bz+c=0\ ,
\end{equation}
where
$
a\equiv 1\ ,
$
$
b\equiv -2\zeta\ ,
$
$
c\equiv 2(\zeta-x)(1-3\zeta^2)/(\zeta x^{2})
$
and $\zeta\equiv x-v$ with $\zeta >0$ and $x>0$ under consideration.
If the two roots of equation (\ref{quadratic1})
are of the same sign
corresponding to the magnetosonic critical point being
a nodal point, the requirement is simply $c/a>0$,
or equivalently $ac>0$;
this translates to
$
(1-3\zeta^{2})(\zeta-x)>0\ .
$
For $x<\sqrt{3}/3\ ,$ we then have
\[
0<-v<\sqrt{3}/3-x\ .
\]
From Figure (\ref{apndxfig}) of the magnetosonic critical
line, we see that this inequality cannot be realized. On
the other hand for $x>\sqrt{3}/3\ ,$ it follows that
\[
\sqrt{3}/3-x<-v<0\ .
\]
Points along the magnetosonic critical line in the range 
of $x>\sqrt{3}/3$ which also satisfy the above inequality 
simultaneously must be in the range of
$
(1+2\lambda)^{1/2}<x<+\infty\ .
$

For the case of two real roots, the discriminant must be
non-negative. The proof is as follows. In the open interval
of $(1+2\lambda)^{1/2}<x<+\infty$, we have the inequality
$\zeta>(1+2\lambda)^{1/2}>1$ and the discriminant is simply
\[
b^{2}-4ac=4\zeta^{2}-8\frac{(\zeta-x)(1-3\zeta^{2})}{x^{2}\zeta}
\]
\begin{equation}
\ \ \quad\qquad =\frac{4x^{2}\zeta^{3}-8(\zeta-x)
(1-3\zeta^{2})}{x^{2}\zeta}\ .
\end{equation}
For a positive denominator, we only need
to examine the sign of the numerator
\[
4x^{2}\zeta^{3}-8(\zeta-x)(1-3\zeta^{2})
\]
\[
=4\zeta^{3}x^{2}+(8-24\zeta^{2})x-(8-24\zeta^{2})\zeta\ .
\]
This quadratic expression in terms of
$x$ approaches the minimum value of
\[
4(-1+6\zeta^{2}-11\zeta^{4}+6\zeta^{6})/\zeta^{3}
\]
as $x$ takes on the value of
$
x=(3\zeta^{2}-1)/\zeta^{3}\ ,
$
where the numerator can be further factorized into
\[
4(3\zeta^{2}-1)(2\zeta^{2}-1)(\zeta+1)(\zeta-1)\ .
\]
For $\zeta>1$, this minimum remains always positive and thus
the discriminant of the critical quadratic equation remains
always positive. We thus come to the conclusion that when the 
abscissa $x$ of a magnetosonic critical point falls within 
the open interval $(1+2\lambda)^{1/2}<x<+\infty$, it is a 
nodal point.

In the case of two real roots with opposite signs (i.e.,
a positive $b^{2}-4ac$ with $ac<0$) for a magnetosonic
critical point being a saddle point, we have
$
(1-3\zeta^{2})(\zeta-x)<0\ .
$
When $x<\sqrt{3}/3$, we have either
\[
-x<-v<0\qquad\hbox{ or }\qquad\sqrt{3}/3-x<-v\ .
\]
As seen from Figure (\ref{apndxfig}),
the latter is actually realizable.

On the other hand when $x>\sqrt{3}/3$, we have either
\[
-x<-v<\sqrt{3}/3-x\qquad\hbox{ or }\qquad 0<-v\ .
\]
Again from Figure (\ref{apndxfig}), the latter can be fulfilled
along the magnetosonic critical line. Also by Figure (\ref{apndxfig}),
it is clear that the magnetosonic critical line always lies on the
upper-right side of the straight line $-v=\sqrt{3}/3-x$, indicating
$-v>\sqrt{3}/3-x$ along the magnetosonic critical line.
Combining these two analyzed cases for saddle type critical points,
we come to the conclusion that for $0<x<(1+2\lambda)^{1/2}$ along
the magnetosonic  critical line, the critical point is of saddle
type.

We now summarize the key results of this Appendix 
A below. Along the magnetosonic critical line, for
\[
0<x<(1+2\lambda)^{1/2}\ ,
\]
the critical point is a saddle point, while for
\[
(1+2\lambda)^{1/2}<x<+\infty\ ,
\]
the critical point is a nodal point.

\begin{figure}
 \includegraphics[width=3in]{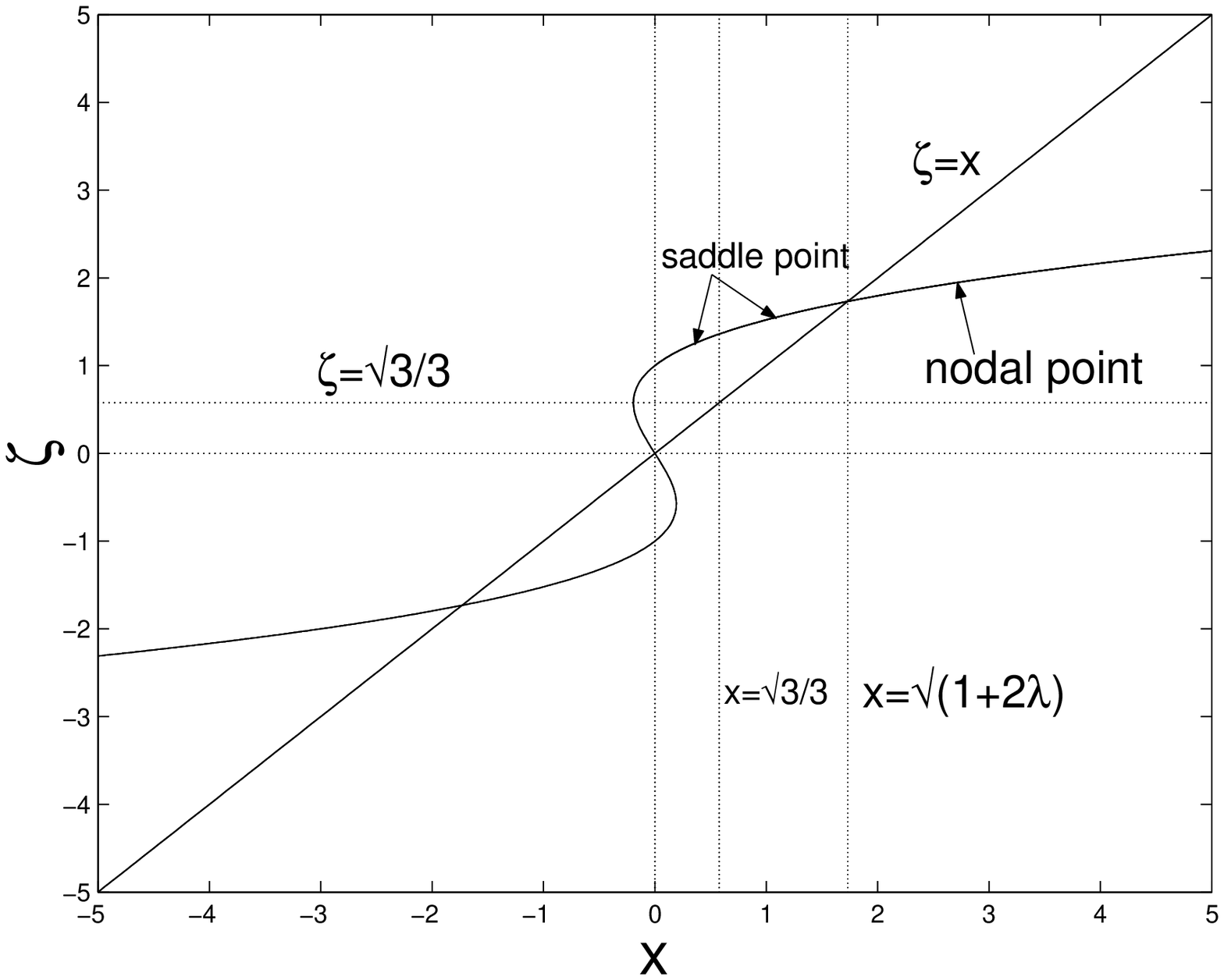}
 \caption{\label{apndxfig} Magnetic field parameter $\lambda=1$, 
 the vertical coordinate is $\zeta\equiv x-v$, and the horizontal 
 coordinate is $x\equiv r/(at)$. The $S$ shaped curve is the 
 magnetosonic critical line expressed in terms of variable 
 $\zeta$. Combining this figure with our analysis, we can
 readily distinguish nodal and saddle points along the 
 magnetosonic critical line. }
\end{figure}

\section[]{Derivation of the Quadratic
Equation for $v^{'}\equiv z$}
Equation (\ref{mainv}) can be rearranged into the form of
\begin{equation}
v^{'}=B/A\ ,
\end{equation}
where
\[
A\equiv (x-v)^{2}-(1+\lambda\alpha x^{2})\ ,
\]
\[
B\equiv (x-v)[\alpha(x-v)-2/x]\ .
\]
Along the magnetosonic critical line, both the denominator
$A$ and the numerator $B$ of the above equation vanish.
We use the L'H\^{o}spital rule to determine the derivatives
at points along the magnetosonic critical line.

The derivative of $A$ with respect $x$ is given by
\[
A'=2(x-v)(1-v^{'})-2\lambda\alpha x-\lambda x^{2}\alpha^{'}\ .
\]
From equation (\ref{alphav}), it follows that
\[
\alpha^{'}=\bigg[\frac{2}{x}(x-v)-v^{'}\bigg]\frac{\alpha}{(v-x)}\ .
\]
Using this expression in the first derivative of $A$, we know that
$A^{'}$ is a linear expression in terms of $v^{'}\equiv z$, namely
\[
A^{'}=f(z)\ ,
\]
where
\[
f(z)\equiv\bigg[\frac{\lambda\alpha x^{2}}
{(v-x)}-2(x-v)\bigg]z+2(x-v)\ .
\]
Meanwhile, we can express the first derivative of $B$ as
\[
B^{'}=g(z)\ ,
\]
where
\[
g(z)\equiv -\frac{2 v[1-x(x - v)\alpha]}{x^{2}}
-\frac{x[-2+x(x - v)\alpha]}{x^{2}}z\ .
\]
Along the magnetosonic critical line, we have the critical
condition that $x(x-v)\alpha=2$. With this relation, $f(z)$
and $g(z)$ can be reduced to the forms of  
\[
f(z)=\bigg[-\frac{2\lambda x}{(x-v)^{2}}-2(x-v)\bigg]z+2(x-v)\ ,
\]
\[
g(z)=2v/x^{2}\ .
\]
By the L'H\^{o}spital rule, we know that
$
z={g(z)}/{f(z)}\ ,
$
or equivalently
$
f(z)z-g(z)=0\ ,
$
leading to quadratic equation (\ref{criticalderivative}).

Note that when $\lambda=0$ and thus $x-v=1$,
we recover the results of Shu (1977), namely
\begin{equation}\label{shueqn}
-2z^{2}+2z-{2(x-1)}/{x^{2}}=0\ .
\end{equation}
The two roots of this equation are simply
\[
z_1={1}/{x_{\ast}}\
\qquad\hbox{ and }\qquad
z_2=1-{1}/{x_{\ast}}\ ,
\]
respectively, where $x_{\ast}$ is a point along the sonic
critical line. For higher-order expansion terms, see more
details of Appendix of Shu (1977) and of Lou \& Shen (2004).

\section[]{Magnetic Pressure and Magnetic Tension Forces}

In spherical polar coordinates ($r,\theta,\phi$), the 
three $r-$, $\theta-$ and $\phi-$components of the 
magnetic tension force are:
\[
(\mathbf{B}\cdot\nabla\mathbf{B})_{r}=
B_{r}\frac{\partial B_{r}}{\partial r}
+\frac{B_{\theta}}{r}\frac{\partial B_{r}}
{\partial\theta}
\]
\begin{equation}\label{tr}
\quad\qquad\qquad\qquad\qquad
+\frac{B_{\phi}}{r\sin\theta}
\frac{\partial B_{r}}{\partial\phi}
-\frac{B^{2}_{\theta}+B^{2}_{\phi}}{r}\ ,
\end{equation}
\[
(\mathbf{B}\cdot\nabla\mathbf{B})_{\theta}=B_{r}\frac{\partial
B_{\theta}}{\partial r}
+\frac{B_{\theta}}{r}\frac{\partial
B_{\theta}}{\partial\theta}
\]
\begin{equation}\label{tt}
\qquad\qquad\qquad
+\frac{B_{\phi}}{r\sin\theta}\frac{\partial
B_{\theta}}{\partial\phi}+\frac{B_{\theta}B_{r}}{r}
-\frac{\cot\theta B^{2}_{\phi}}{r}\ ,
\end{equation}
\[
(\mathbf{B}\cdot\nabla\mathbf{B})_{\phi}
=B_{r}\frac{\partial B_{\phi}}{\partial r}
+\frac{B_{\theta}}{r}\frac{\partial
B_{\phi}}{\partial\theta}
\]
\begin{equation}\label{tp}
\qquad\qquad\qquad
+\frac{B_{\phi}}{r\sin\theta}\frac{\partial
B_{\phi}}{\partial \phi}+\frac{B_{\phi}B_{r}}{r}
+\frac{\cot\theta B_{\phi}B_{\theta}}{r}\ .
\end{equation}
Meanwhile, the radial component of 
the magnetic pressure gradient is
\[
\bigg[\frac{1}{2}\nabla({\mathbf B}^{2})\bigg]_{r}
=\frac{1}{2}\frac{\partial}{\partial r}
(B^{2}_{r}+B^{2}_{\theta}+B^{2}_{\phi})
\]
\[
\quad\quad\quad=B_{r}\frac{\partial B_{r}}{\partial r}
+\frac{1}{2}\frac{\partial}{\partial r}(B^{2}_{\theta}
+B^{2}_{\phi})
\]
\begin{equation}\label{gr}
\quad\quad\quad=B_{r}\frac{\partial B_{r}}{\partial r}
+B_{\theta}\frac{\partial B_{\theta}}{\partial r}
+B_{\phi}\frac{\partial B_{\phi}}{\partial r}\ ;
\end{equation}
the $\theta$ component of the magnetic pressure gradient is
\[
\bigg[\frac{1}{2}\nabla({\mathbf B}^{2})\bigg]_{\theta}
=\frac{1}{2 r}\frac{\partial}{\partial\theta}
(B^{2}_{r}+B^{2}_{\theta}+B^{2}_{\phi})
\]
\begin{equation}\label{gt}
\quad\quad\quad=\frac{B_{r}}{r}\frac{\partial B_{r}}
{\partial\theta}
+\frac{B_{\theta}}{r}\frac{\partial B_{\theta}}
{\partial\theta}
+\frac{B_{\phi}}{r}\frac{\partial B_{\phi}}
{\partial\theta}\ ;
\end{equation}
and the $\phi$ component of the magnetic pressure gradient is
\[
\bigg[\frac{1}{2}\nabla({\mathbf B}^{2})\bigg]_{\phi}
=\frac{1}{2 r\sin\theta}\frac{\partial}{\partial\phi}
(B^{2}_{r}+B^{2}_{\theta}+B^{2}_{\phi})
\]
\begin{equation}\label{gp}
\quad\quad\quad=\frac{B_{r}}{r\sin\theta}\frac{\partial
B_{r}}{\partial \phi}+\frac{B_{\theta}}{r\sin\theta}
\frac{\partial B_{\theta}}{\partial\phi}
+\frac{B_{\phi}}{r\sin\theta}
\frac{\partial B_{\phi}}{\partial\phi}\ .
\end{equation}
Strictly speaking, when the magnetic field is taken into account,
a global spherical symmetry cannot be sustained. Checking through
equations (\ref{tr}), (\ref{tt}), (\ref{tp}), and equations (\ref{gr}),
(\ref{gt}), (\ref{gp}), we see that second and third terms in equation 
(\ref{tr}) are ignored. The first term in (\ref{tr}) and the first 
term in (\ref{gr}) cancel each other. In the radial momentum equation, 
the fourth term in equation (\ref{tr}) and the second and third terms 
in equation (\ref{gr}) are present and kept.

As we take the magnetofluid system to be quasi-spherically symmetric
on larger scales, the terms involving $\partial/\partial\theta$ and
$\partial/\partial\phi$ are neglected. Thus in the $\theta$ and
$\phi$ directions, the first, fourth and fifth terms in equations
(\ref{tt}) and (\ref{tp}) would cause the overall magnetofluid to
be nonspherically symmetric. In order to work in the framework of
a quasi-spherical symmetry, we would drop these these terms. 
Nevertheless, we should keep in mind that these terms might give 
rise to nonspherical shapes of various nebulae in contexts of
astrophysics. The quasi-elliptical morphology of the Crab Nebula
is a concrete example. 

\label{lastpage}


\vskip 0.4cm

\begin{thebibliography}{99}
\bibitem[\protect\citeauthoryear{Adams, Lada \& Shu}{1987}]{als}
Adams F. C., Lada C. J.,  Shu F. H., 1987, ApJ, 321, 788

\bibitem[\protect\citeauthoryear{Balick \& Frank}{2002}]{balick02}
Balick B., Frank A., 2002, ARA\&A, 40, 439

\bibitem[\protect\citeauthoryear{Bertschinger}{1985}]{bertschinger85}
Bertschinger E., 1985, ApJS, 58, 39

\bibitem[\protect\citeauthoryear{BianLou}{2005}]{BianLou05}
Bian F.-Y., Lou Y.-Q., 2005, MNRAS, submitted

\bibitem[\protect\citeauthoryear{Blackman et al.}{2001}]{blackman01}
Blackman E. G., Frank A., Markiel J. A., Thomas J. H., Van Horn H. M.,
2001, Nature, 409, 485

\bibitem[\protect\citeauthoryear{Bodenheimer \&
Sweigart}{1968}]{bs68} Bodenheimer P., Sweigart A., 1968, 152, 515

\bibitem[\protect\citeauthoryear{Bondi}{1952}]{bondi52} Bondi H.,
1952, MNRAS, 112, 195

\bibitem[\protect\citeauthoryear{Boily \&
Lynden-Bell}{1995}]{boily95} Boily C. M., Lynden-Bell D., 1995,
MNRAS, 276, 133

\bibitem[\protect\citeauthoryear{Bonnor}{1956}]{bonnor56} Bonnor
W. B., 1956, MNRAS, 116, 351

\bibitem[\protect\citeauthoryear{Bouquet et al.}{1985}]{bouquet85} Bouquet
S., Feix M. R., Fijalkow E., Munier A., 1985, ApJ, 293, 494

\bibitem[\protect\citeauthoryear{CaiShu}{2005}]{caishu05}Cai M. J.,
Shu, F. H., 2005, ApJ, 618, 438

\bibitem[\protect\citeauthoryear{Chandrasekhar}{1957}]
{chandrasekhar57}Chandrasekhar S., 1957, Stellar
Structure. Dover Publications, New York

\bibitem[\protect\citeauthoryear{Cheng}{1978}]{cheng78}
Cheng A. F., 1978, ApJ, 221, 320

\bibitem[\protect\citeauthoryear{Chiueh \& Chou}{1994}]{chiueh}
Chiueh T. H., Chou J. K., 1994, ApJ, 431, 380

\bibitem[\protect\citeauthoryear{Choi et al.}{1995}]{choi95} Choi
M., Evans N. J., II, Gregersen E. M., Wang Y., 1995, ApJ, 448, 742

\bibitem[\protect\citeauthoryear{Contopoulos et
al.}{1998}]{contopoulos98} Contopoulos I., Ciolek G. E.,
K\"{o}nigl A., 1998, ApJ, 504, 247

\bibitem[\protect\citeauthoryear{Ebert}{1955}]{ebert55} Ebert R.,
Zs. Ap., 37, 217

\bibitem[\protect\citeauthoryear{Fabian}{1994}]{fabian94} Fabian
A. C., 1994, ARA\&A, 32, 277

\bibitem[\protect\citeauthoryear{Falgarone et
al.}{2001}]{falgarone01} Falgarone E., Pety J., Phillips T. G.,
2001, ApJ, 555, 178

\bibitem[\protect\citeauthoryear{Fillmore \&
Goldreich}{1984}]{fillmore84} Fillmore J. M., Goldreich P., 1984,
ApJ, 284, 1

\bibitem[\protect\citeauthoryear{Foster \&
Chevalier}{1993}]{fosterchevalier93} Foster P. N., Chevalier R.
A., 1993, ApJ, 416, 303

\bibitem[\protect\citeauthoryear{Goldreich \&
Weber}{1980}]{goldreich80} Goldreich P., Weber S. V., 1980, 238,
991

\bibitem[\protect\citeauthoryear{Gunn \& Gott}{1972}]{gunn72} Gunn
J. E., Gott III J. R., 1972, ApJ, 176, 1

\bibitem[\protect\citeauthoryear{Hanawa}{1997}]{hanawa97} Hanawa
T., Nakayama K., 1997, ApJ, 484, 238

\bibitem[\protect\citeauthoryear{Hanawa \&
Matsumoto}{1999}]{hanawa99} Hanawa T., Matsumoto T., 1999, ApJ,
521, 703

\bibitem[\protect\citeauthoryear{Hanawa \&
Matsumoto}{2000}]{hanawa00} Hanawa T., Matsumoto T., 2000, PASJ,
52, 241

\bibitem[\protect\citeauthoryear{Harada et al.}{2003}]{harada03}
Harada T., Maeda H., Semelin B., 2003, Phys. Rev. D, 67, 084003

\bibitem[\protect\citeauthoryear{Hartmann}{1998}]{hartmann}
Hartmann L., 1998, Accretion Processes in Star Formation,
Cambridge University Press

\bibitem[\protect\citeauthoryear{Harvey et al.}{2001}]{harvey01}
Harvey D. W., Wilner D. J., Myers P. C., Alves J. F., Chen H.,
2001, ApJ, 563, 903

\bibitem[\protect\citeauthoryear{Hayashi}{1966}]{hayashi66}
Hayashi C., 1966, ARA\&A, 4, 171

\bibitem[\protect\citeauthoryear{Hennebelle}{2003}]{hennebelle03}
Hennebelle P., 2003, A\&A, 397, 381

\bibitem[\protect\citeauthoryear{Hunter}{1977}]{hunter77} Hunter
C., 1977, ApJ, 218, 834

\bibitem[\protect\citeauthoryear{Hunter}{1986}]{hunter86} Hunter
C., 1986, MNRAS, 223, 391

\bibitem[\protect\citeauthoryear{Inutsuka \&
Miyama}{1992}]{inutsuka92} Inutsuka S.,
Miyama S. M., 1992, ApJ, 388, 392

\bibitem[\protect\citeauthoryear{Jordan \& Smith}{1977}]{jordan77}
Jordan D. W., Smith P., 1977, Nonlinear Ordinary Differential
Equations. Oxford Univ. Press, Oxford

\bibitem[\protect\citeauthoryear{Kawachi
\& Hanawa}{1998}]{kawachi98} Kawachi T.,
Hanawa T., 1998, PASJ, 50, 577

\bibitem[\protect\citeauthoryear{Klessen \&
Burkert}{2000}]{klessen00} Klessen R. S., Burkert A., 2000, ApJS,
128, 287

\bibitem[\protect\citeauthoryear{Krasnopolsky \&
K\"{o}nigl}{2002}]{krasnopolsky02} Krasnopolsky R., K\"{o}nigl A.,
2002, ApJ, 580, 987

\bibitem[\protect\citeauthoryear{Kwok}{1982}]{kwok82} Kwok S.,
1982, ApJ, 258, 280

\bibitem[\protect\citeauthoryear{Kwok}{1985}]{kwok85} Kwok S.,
1985, AJ, 90, 49

\bibitem[\protect\citeauthoryear{Kwok}{1993}]{kwok93} Kwok S.,
1993, ARA\&A, 31, 63

\bibitem[\protect\citeauthoryear{Lai}{2000}]{lai00} Lai D., 2000,
ApJ, 540, 946

\bibitem[\protect\citeauthoryear{Lai \&
Goldreich}{2000}]{laigoldreich00} Lai D., Goldreich P., 2000, ApJ,
535, 402

\bibitem[\protect\citeauthoryear{Landau \&
Lifshitz}{1959}]{landau59} Landau L. D., Lifshitz E. M.,
1959, Fluid Mechanics (New York: Pergamon)

\bibitem[\protect\citeauthoryear{Larson}{1969a}]{larson69a} Larson
R. B., 1969a, MNRAS, 145, 271

\bibitem[\protect\citeauthoryear{Larson}{1969b}]{larson69b}
Larson R. B., 1969b, MNRAS, 145, 405

\bibitem[\protect\citeauthoryear{Lou}{1993}]{lou93}
Lou Y. Q., 1993, ApJ, 414, 656

\bibitem[\protect\citeauthoryear{Lou}{1994}]{lou94}
Lou Y. Q., 1994, ApJ, 428, L21

\bibitem[\protect\citeauthoryear{Lou \& Shen}{2004}]{loushen}
Lou Y. Q., Shen Y., 2004, MNRAS, 348, 717

\bibitem[\protect\citeauthoryear{Maeda et al.}{2002}]{maeda02}
Maeda H., Harada T., Iguchi H., Okuyama N., 2002, Progress of
Theoretical Physics, 108, 819

\bibitem[\protect\citeauthoryear{McLaughlin \&
Pudritz}{1997}]{mclaughlin97} McLaughlin D. E., Pudritz R. E.,
1997, ApJ, 476, 750

\bibitem[\protect\citeauthoryear{Murakami, Nishihara \&
Hanawa}{2004}]{murakami04} Murakami M., Nishihara K., Hanawa T.,
2004, ApJ, 607, 879

\bibitem[\protect\citeauthoryear{Navarro et al.}{1997}]{navarro97}
Navarro J. F., Frenk C. S., White S. D. M., 1997, ApJ, 490, 493

\bibitem[\protect\citeauthoryear{Ori \& Piran}{1988}]{ori88} Ori
A., Piran T., 1988, MNRAS, 234, 821

\bibitem[\protect\citeauthoryear{Ostriker et
al.}{2001}]{ostriker01} Ostriker E. C., Stone J. M., Gammie C. F.,
2001, ApJ, 546, 980

\bibitem[\protect\citeauthoryear{Parker}{1963}]{parker63} Parker
E. N., 1963, Interplanetary Dynamical Processes (New York: Wiley)

\bibitem[\protect\citeauthoryear{Penston}{1969a}]{penston69a}
Penston M. V., 1969a, MNRAS, 144, 425

\bibitem[\protect\citeauthoryear{Penston}{1969b}]{penston69b}
Penston M. V., 1969b, MNRAS, 145, 457

\bibitem[\protect\citeauthoryear{Porter et al.}{1994}]{porter94}
Porter D. H., Pouquet A., Woodward P. R., 1994, Phys. Fluid, 6,
2133

\bibitem[\protect\citeauthoryear{Press et al.}{1986}]{press86}
Press W. H., Flannery B. P., Teukolsky S. A., Vetterling W.,
1986, Numerical Recipes (Cambridge: Cambridge Univ. Press)

\bibitem[\protect\citeauthoryear{Sarazin}{1988}]{sarazin88}
Sarazin C. L., 1988, X-Ray Emission from Clusters of Galaxies.
Cambridge University Press, Cambridge

\bibitem[\protect\citeauthoryear{Saito et al.}{1999}]{sa99} Saito
M., Sunada K., Kawabe R., Kitamura Y., Hirano N., 1999, ApJ, 518,
334

\bibitem[\protect\citeauthoryear{Scalo et al.}{1998}]{scalo98}
Scalo J., Vazquez-Semadeni E., Chappell D., Passot T., 1998, ApJ,
504, 835

\bibitem[\protect\citeauthoryear{Sedov}{1959}]{sedov59} Sedov L.
I., 1959, Similarity and Dimensional Methods in Mechanics (New
York: Academic)

\bibitem[\protect\citeauthoryear{Semelin, Sanchez \& de
Vega}{2001}]{semelin01} Semelin B., Sanchez N., de Vega H. J.,
2001, Phys. Rev. D, 63, 4005

\bibitem[\protect\citeauthoryear{Shadmehri}{2005}]{shadmehri05}
Shadmehri M., 2005, 356, 1429

\bibitem[\protect\citeauthoryear{Shen \& Lou}{2004}]{shenlou04}
Shen Y., Lou Y. Q., 2004, ApJL, 611, 117

\bibitem[\protect\citeauthoryear{Shu}{1977}]{shu77} Shu F. H.,
1977, ApJ, 214, 488

\bibitem[\protect\citeauthoryear{Shu, Adams \&
Lizano}{1987}]{sal87} Shu F. H., Adams F. C.,
Lizano S., 1987, ARA\&A, 25, 23

\bibitem[\protect\citeauthoryear{Shu et al.}{2002}]{shu02}
Shu F. H., Lizano S., Galli D., Cant\'o J., Laughlin G.,
2002, ApJ, 580, 969

\bibitem[\protect\citeauthoryear{Spitzer}{1978}]{spitzer78}
Spitzer L., Physical Processes in the Interstellar Medium.
Wiley, New York

\bibitem[\protect\citeauthoryear{Suto \& Silk}{1988}]
{suto88}Suto Y., Silk J., 1988, ApJ, 326, 527

\bibitem[\protect\citeauthoryear{Tanaka \& Washimi}{2002}]
{tanaka02}Tanaka T., Washimi H., 2002, Science, 296, 321

\bibitem[\protect\citeauthoryear{Terebey, Shu \& Cassen}{1984}]
{tsc}Terebey S., Shu F. H., Cassen P., 1984, ApJ, 286, 529

\bibitem[\protect\citeauthoryear{Tilley \& Pudritz}{2003}]
{tp03}Tilley D. A., Pudritz R. E., 2003, ApJ, 593, 426

\bibitem[\protect\citeauthoryear{Tsai \& Hsu}{1995}]{th95}
Tsai J. C., Hsu J. J. L., 1995, ApJ, 448, 774

\bibitem[\protect\citeauthoryear{Whitworth \&
Summers}{1985}]{whitworth85} Whitworth A.,
Summers D., 1985, MNRAS, 214, 1

\bibitem[\protect\citeauthoryear{Wolf et al.}{2003}]{wolf03}
Wolf S., Launhardt R., Henning T., 2003, ApJ, 592, 233

\bibitem[\protect\citeauthoryear{Yahil}{1983}]{yahil83}
Yahil A., 1983, ApJ, 265, 1047

\bibitem[\protect\citeauthoryear{Zhou et al.}{1993}]{zhou93}Zhou
S., Evans N. J., II, K\"ompe C., Walmsley C. M., 1993, ApJ, 404, 232

\end{thebibliography}
\end{document}